\documentclass[aps,prd,onecolumn,showpacs,showkeys,amsmath,amssymb]{revtex4}
\usepackage{graphicx}
\usepackage{dcolumn}
\usepackage{bm}
\usepackage[dvips]{color}
\include{shorthand}
\usepackage{amssymb}
\usepackage{amsmath}

\def\half{\frac{1}{2}}

\def\inner#1#2{{\bm #1}\cdot {\bm #2}}

\def\half{\frac{1}{2}}
\def\half{\frac{1}{2}}

\begin{document}

\title{Baryon fields with $U_{L}(3)\times U_{R}(3)$ chiral symmetry V: \\
Pion-Nucleon and Kaon-Nucleon $\Sigma$ Terms}
\author{V. Dmitra\v sinovi\' c$^1$}
\email{dmitrasin@ipb.ac.rs}
\author{Hua-Xing Chen$^{2}$}
\email{hxchen@buaa.edu.cn}
\author{Atsushi Hosaka$^{3}$}
\email{hosaka@rcnp.osaka-u.ac.jp}
\affiliation{$^1$Institute of Physics, Belgrade University, Pregrevica 118, Zemun, P.O.Box 57, 11080 Beograd, Serbia
\\
$^2$ School of Physics and Nuclear Energy Engineering and International Research
Center for Nuclei and Particles in the Cosmos, Beihang University, Beijing 100191, China
\\
$^3$ Research Center for Nuclear Physics, Osaka University, Ibaraki 567--0047, Japan}
%
\begin{abstract}
We have previously calculated the pion-nucleon $\Sigma_{\pi N}$ term in the chiral mixing approach with
$u,d$ flavors only, and found the lower bound
$\Sigma_{\pi N} \geq \left(1 + \frac{16}{3} \sin^2 \theta \right){3 \over 2}
\left(m_{u}^{0} + m_{d}^{0}\right)$
where $m_{u}^{0}, m_{d}^{0}$ are the current quark masses, and $\theta$
is the mixing angle of the
$[(\mathbf{\frac12},\mathbf{0})\oplus (\mathbf{0},\mathbf{\frac12})]$
and the $[(\mathbf{1},\mathbf{\frac12})\oplus (\mathbf{\frac12},\mathbf{1})]$
chiral multiplets.
This mixing angle can be calculated as $\sin^2\theta = \frac{3}{8}\left(g_A^{(0)} + g_A^{(3)}\right)$,
where $g_A^{(0)}, g_A^{(3)}$, are the flavor-singlet and the isovector axial couplings.
With presently accepted values of current quark masses,
this leads to $\Sigma_{\pi N} \geq 58.0 \pm 4.5 \begin{array}{l} +11.4 \\ -6.5 \end{array}$
MeV, which is in agreement with the values extracted from experiments, and substantially
higher than most previous two-flavour calculations.
The causes of this enhancement are: 1) the large, ($\frac{16}{3}\simeq 5.3$),
purely $SU_L(2) \times SU_R(2)$ algebraic factor; 2) the admixture of the
$[(\mathbf{1},\mathbf{\frac12})\oplus (\mathbf{\frac12},\mathbf{1})]$ chiral multiplet
component in the nucleon, whose presence has been known for some time, but that had not
been properly taken into account, yet.
We have now extended these calculations of $\Sigma_{\pi N}$ to three light flavours,
i.e., to $SU_L(3) \times SU_R(3)$ multiplet mixing.
Phenomenology of chiral $SU_L(3) \times SU_R(3)$ multiplet mixing demands the presence of
three chiral $SU_L(3) \times SU_R(3)$ multiplets, {\it viz.}
$[(\mathbf{6},\mathbf{3})\oplus (\mathbf{3},\mathbf{6})]$,
$[(\mathbf{3},\bar{\mathbf{3}})\oplus(\bar{\mathbf{3}},\mathbf{3})]$ and
$[(\bar{\mathbf{3}},\mathbf{3})\oplus(\mathbf{3},\bar{\mathbf{3}})]$, in order to
successfully reproduce the baryons' flavor-octet and flavor-singlet axial currents,
as well as the baryon anomalous magnetic moments. Here we use these previously obtained
results, together with known 
constraints on the explicit chiral symmetry breaking in baryons
to calculate the $\Sigma_{\pi N}$ term, 
but find little, or no change of $\Sigma_{\pi N}$ from the above successful two-flavor result.
The physical significance of these results lies in the fact that they show no need for
$q^4 {\bar q}$ components, and in particular, no need for an $s \bar s$ component in the
nucleon, 
in order to explain the large ``observed'' $\Sigma_{\pi N}$ value. We also predict the
experimentally unknown kaon-nucleon sigma term $\Sigma_{K N}$.
\end{abstract}
\maketitle
\pagenumbering{arabic}
%

\section{Introduction}

For more than 35 years the deviation of the nucleon $\Sigma_{\pi N}$ term
extracted from the measured $\pi N$ scattering partial wave analyses
(in the following to be called ``measured value'', for brevity) from the
(naive quark model) value of 25 MeV
was interpreted as an increase of Zweig-rule-breaking in the nucleon, or
equivalently to an increased content of (unpolarized) $s \bar s$ pairs in the
nucleon~\cite{Cheng:1975wm,Donoghue:1985bu,kj87}, defined as
$y = \frac{2 \langle N |{\bar s} s | N \rangle}{\langle N |{\bar u} u + {\bar u} u| N \rangle}$.
Moreover, the anomalously small measured value of the flavor-singlet axial coupling
$g_A^{(0)} = 0.33 \pm 0.06$~\cite{Alexakhin:2006oza,Airapetian:2006vy,Ageev:2007du},
or the older value $0.28 \pm 0.16$~\cite{Filippone:2001ux}, as compared with the naive
quark model prediction of $g_A^{(0)} = 1$, was long interpreted as 
evidence for an increased {\it polarized} $s \bar s$ content of
the nucleon~\cite{Ratcliffe:1990dh,Filippone:2001ux,Vogelsang:2007zza,Shore:2007yn}.
Yet more recently, both of these conclusions and interpretations 
were checked directly in low-momentum transfer $Q$ parity-violating elastic electron
scattering experiments, however, and were found to be incorrect~\cite{Acha:2006my,Baunack:2009gy,Androic:2009aa}.

Whereas this situation is consistent with QCD, it seems in contradiction with earlier
expectations, that were based
on a combination of quark and chiral effective field theory models~\cite{Cheng:1985bj,Donoghue:1992dd,Bass:2007zzb}. The question remains if one
can explicitly construct an effective chiral field theory model for nucleons and
mesons that is
connected to the underlying quark structure of hadrons
and reproduces these two ``anomalous'' results?

Gell-Mann and L\' evy's (GML) linear sigma model has been the principal
example
of an effective field theory model of strongly interacting nucleons
and pions with spontaneously broken chiral symmetry ever since its inception more than 50 years ago~\cite{GellMann:1960np,Bernstein:1960}.
It is well known that this linear sigma model does not always reproduce
the correct phenomenology,
e.g. a) the value of
the isovector axial coupling strength $g_A^{(1)}$ equals unity in this model;
b) the value of the isoscalar pion-nucleon scattering length is too large in this model.

Both of these shortcomings have been removed in an extended linear sigma model, proposed by
Bjorken and Nauenberg \cite{Bjorken:1968ej} and by Lee \cite{lee72}:
a) The first one had been fixed by introducing an additional derivative-coupling term that is
not renormalizable. 
b) Fred Myhrer and one of us (V.D.) showed 
in Ref. \cite{Dmitrasinovic:1999mf} that consequently
the phenomenology is considerably improved in the Bjorken-Nauenberg-Lee (BNL) extended linear
sigma model, as compared to the original GML model; in particular, the value of the isoscalar
pion-nucleon scattering length is reduced to its observed value.
This improvement is directly related to the correct value of the (isovector) nucleon axial
coupling constant $g_A^{(1)}$ in the 
BNL model, 
which in turn is a direct consequence of the new derivative coupling~\footnote{This BNL extended linear sigma model allows one to study the $g_A^{(1)}$ dependence of the
$\pi$N scattering lengths, $a_{\pi N}$,  and of the nucleon sigma term $\Sigma_{\pi N}$.
It is well known that $a_{\pi N}^{(-)}$ depends crucially on the value of $g_A^{(1)}$,
whereas the  $\Sigma_{\pi N}$ dependence on $g_A^{(1)}$ was not known (and may be model-dependent).
We displayed this dependence of $a_{\pi N}^{(-)}$ and showed that a large value of
$\Sigma_{\pi N}$ could easily be obtained without recourse to any $s\bar{s}$ component of the
nucleon with the values of bare parameters available at the time (which have changed
drastically in the meantime, however). We also reproduced the then new,
tiny experimental value of the isoscalar $\pi$N scattering length $a_{0}^{(+)}$.}.
This shows the phenomenological importance of having the correct value of the isovector
axial coupling.

That is not the only axial coupling of the nucleon, however: there
is also the isoscalar one $g_A^{(0)}$, whose measured value $g_A^{(0)} = 0.33 \pm 0.08$, or $0.28 \pm 0.16$
deviates even more from unity, which is the value that the naive non-relativistic quark model suggests
and the GML model postulates.
The BNL derivative coupling term does not fix the value of the isoscalar axial coupling strength
$g_A^{(0)}$, however. Here one could continue with the BNL stratagem
and introduce yet another derivative-coupling term to fix this problem, but clearly that would be
{\it ad hoc} and in no apparent way related to the underlying quark structure.

An alternative approach was attempted with the notion of chiral representation mixing,
which is in fact older than the BNL model, but was able to reproduce realistic values of the isovector
axial couplings~\cite{Harari:1966jz,Harari:1966yq,Gerstein:1966zz}.
By choosing low dimensional representations, as corresponding to the ones of the nucleon's three-quark
interpolators at present days, that approach turned out to give some constraints on the values of the axial couplings $g_A^{(0,1)}$
without introducing derivative couplings of hadrons~\cite{Nagata:2007di,Nagata:2008xf,Nagata:2008zzc}.
In this sense it is rather different from the 
BNL model, and one should not be surprised if other predictions of the two models are different.
What is perhaps not so well known is that there are two-flavor (linear realization) sigma model chiral
Lagrangians
based on the concept of chiral mixing, that reproduce the two-flavor chiral low-energy theorems~\cite{Hara:1965,lee72,Christos,Nagata:2008zzc,Nagata:2008xf,Dmitrasinovic:2009vy,Gallas:2013ipa}.
Over the years, these Lagrangians have been extended to three flavors~\cite{Christos2,Zheng:1991pk,Zheng:1992mn,Chen:2009sf,Chen:2010ba,Chen:2011rh,Nishihara:2015fka}
and adapted/fitted to the two axial couplings and other nucleon properties, such as the magnetic moments
\cite{Love:1970vi,Dmitrasinovic:2011yf}.

An advantage of the chiral representation mixing is that the possible representations
and their mixing may be inferred by the quark structure of the nucleon.
For instance, in the Schwinger-Dyson-Faddeev-Bethe-Salpeter approach to
QCD \cite{Eichmann:2009qa}, different Dirac structures in the Faddeev-Bethe-Salpeter equation
are sources of different chiral representation (or components in the Faddeev-Bethe-Salpeter amplitude),
thus leading to a mixing
of chiral representations in the physical nucleon wave function.


The purpose of the present paper is to coherently and systematically present the calculation
of the pion-nucleon sigma term $\Sigma_{\pi N}$ and of the isoscalar axial coupling $g_A^{(0,1)}$
in the chiral-mixing linear sigma model.
The isovector axial coupling has been studied by many authors as already mentioned above.
The isoscalar axial coupling $g_A^{(0)}$ had been calculated in
Refs. \cite{Dmitrasinovic:2009vp,Dmitrasinovic:2009vy}, and in
Refs. \cite{Dmitrasinovic:2014aya,Dmitrasinovic:2014eia} we have briefly presented
our results for the pion-nucleon sigma term $\Sigma_{\pi N}$ in the chiral mixing approach.
The result of $\Sigma_{\pi N}$
depends substantially on $g_A^{(1)}$, in contrast to the BNL model one \cite{Dmitrasinovic:1999mf},
and agrees with the experimental value (almost embarrassingly) well. This phenomenological success
has been a source of some open and more hidden criticism.
We do not wish to over-emphasize this phenomenological agreement here, as it is subject to the
time-dependent variation of the free parameters, more specifically, to the current quark masses, which were
about 50\%  larger 15 years ago - in this model, but rather we try and systematically explore the differences
among various effective chiral models.


Moreover, we make a systematic exposition of our approach and we record this model's
predictions of the kaon-nucleon sigma term, which have not been measured as yet,
just in case some day they are measured if only on the lattice, and thus open ourselves
to potential future criticism.
In this way we explicitly show how to construct effective linear chiral model(s) of interacting
nucleons and mesons based on the underlying quark structure, that does not need $s \bar s$ content in the
nucleon to reproduce the two crucial observables, the $\Sigma_{\pi N}$ and the $g_A^{(0)}$.

The crucial assumption here is the systematic implementation of chiral symmetry at
all levels, i.e., at both the quark and the hadron levels.
In Ref.~\cite{Dmitrasinovic:1999mf} we have shown how Weinberg's ``chiral boost'' transformation
of the BNL model leads to a nonlinear realization chiral Lagrangian,
as corresponding to the leading order of the chiral perturbation theory Lagrangian.
The same procedure can be applied to the chiral mixing Lagrangian, with the same result.
For this to happen, two different linear Lagrangians lead to the same nonlinear one.
That goes to show that the linear-to-nonlinear-realization mapping (Weinberg's chiral boost)
is of the many-to-one kind. Thus, it ``hides'' many details of the underlying dynamics at low
values of momentum transfer as compared with 
$f_{\pi}$= 93 MeV and $m_{\pi}$= 140 MeV,
and emphasizes the dynamical aspects of chiral symmetry.
In that sense, the nonlinear realization can be viewed as being ``coarser'' than the linear one.
These dynamical details become increasingly visible as the momentum transfer is increased, however.

We believe that at least some of the generally valid chiral predictions of all chiral models
are most economically obtained in the chiral-mixing model. In particular, we believe
that the role of $U_A(1)$ symmetry and its breaking in the baryon sector has been
ignored thus far, and our study appears to be the first step in
rectifying this lamentable situation.

Throughout this paper we shall use the first Born approximation at the tree level.
In order to explore the various possibilities and to facilitate comparison
with earlier studies of the Gell-Mann--Levy linear sigma model, we
introduce three different chiral symmetry breaking
[$\chi$SB] terms, as in Refs.~\cite{camp79,bc79}.

This paper is the fifth one in a sequence of papers~\cite{Chen:2008qv,Chen:2009sf,Chen:2010ba,Chen:2011rh},
consequently, we shall repeat here, for the sake of completeness and coherence of presentation, rather
than merely cite, several (a bare minimum) of equations and tables that have already
appeared in our previous papers.

The paper falls into six sections and two appendixes.
In Section~\ref{sect:Phenomenology} we consider the chiral mixing phenomenology.
Then in Section~\ref{sect:sigma model}, which is devoted to a
construction of a two- and three-flavor chiral Lagrangians that reproduce
the chiral mixing phenomenology, we
present the $\chi$SB terms and the canonical field variables, and
show that the Noether charges close the chiral algebra although $g_A \ne 1$.
In Section~\ref{sect:pi_Sigma_term} we examine the pion-nucleon sigma
term $\Sigma_{\pi N}$ - some of these results have been reported at conferences~\cite{Dmitrasinovic:2014aya,Dmitrasinovic:2014eia}.
%
In Section~\ref{sect:K_Sigma_term} we examine the kaon-nucleon
$\Sigma_{K N}$ term, and in
Section~\ref{sec:summary} we summarize the results.

\section{Phenomenology of chiral mixing}
\label{sect:Phenomenology}

The basic premise of chiral mixing approach is that the chiral $SU_L(3) \times SU_R(3)$
symmetry is spontaneously broken and therefore that the 
eigenstates do not form (ir)reducible representations of the chiral symmetry group
$SU_L(3) \times SU_R(3)$~\footnote{Chiral symmetry does not 
require irreducible representations for parity-conserving interactions.}.
Rather, the 
eigenstates are linear superpositions of several
(ir)reducible representations of $SU_L(3) \times SU_R(3)$.
In general, such chiral representation mixing theories tend to be
most powerful and predictive when only a few chiral multiplets are involved.
As the number of admixed multiplets grows, this method becomes
increasingly complicated and thus loses its predictive power.

Just which irreducible representations are being admixed, is a question that
ultimately ought to be answered by QCD. In the absence of a QCD-based answer,
the choice can be (severely) limited by the following mathematical and 
physical considerations.

\subsection{Chiral representations}

Group-theoretical considerations impose limitations on the allowed
(ir)reducible representation of $SU_L(3) \otimes SU_R(3)$:
any (ir)reducible representation of $SU_L(3) \otimes SU_R(3)$, that is described
by two $SU(3)$ irreducible representations, $(G_L,G_R)$, leads to 
irreducible representations $G_F$ of $SU_F(3)$ as determined by the Clebsch-Gordan
series of the tensor product: $G_F \in G_L \otimes G_R$. For example,
${\bf 10} \oplus {\bf 8} = {\bf 6} \otimes {\bf 3} \in [({\bf 6},{\bf 3})\oplus({\bf 3},{\bf 6})]$,
${\bf 1} \oplus {\bf 8} = \mathbf{\bar 3} \otimes \mathbf{3} \in
[(\mathbf{\bar 3},\mathbf{3})\oplus(\mathbf{3},\mathbf{\bar 3})]$,
and
${\bf 8} = \mathbf{8} \otimes \mathbf{1} \in [(\mathbf{8},\mathbf{1})\oplus(\mathbf{1},\mathbf{8})]$.

If one demands that only the experimentally observed irreducible representations $G$ of
$SU_F(3)$ appear in these Clebsch-Gordan series, then one is limited to the above three
reducible chiral representations: Any other chiral representation, other than the (trivial)
chiral-singlet one, $[(\mathbf{1},\mathbf{1})]$, necessarily leads to $SU_F(3)$-exotics.

When we further take into account the so-called ``mirror'' representations, in which the left- (L) and
the right-handed (R) representations are interchanged, ($G_L \leftrightarrow G_R$), in the chiral
multiplet, then the number of allowed chiral multiplets is six.
Mathematically, there is no difference between the ``naive'' (natural?) and ``mirror'' representations;
physically, and historically, the ``naive'' ones were introduced first, mostly because there were no
explicit examples how the ``mirror'' ones could arise in a three-quark system. That ``objection'' was
finally raised by explicit examples of mirror (three-quark) interpolating fields in
Refs. \cite{Nagata:2007di,Nagata:2008xf,Nagata:2008zzc,Chen:2008qv,Chen:2012ex,Chen:2012vs}.
For octet baryons, this limits the permissible chiral multiplets to
$[({\bf 6},{\bf 3})\oplus({\bf 3},{\bf 6})]$,
and its ``mirror'' $[({\bf 3},{\bf 6})\oplus({\bf 6},{\bf 3})]$, to
$[(\mathbf{\bar 3},\mathbf{3})\oplus(\mathbf{3},\mathbf{\bar 3})]$, and its ``mirror''
$[(\mathbf{3},\mathbf{\bar 3})\oplus(\mathbf{\bar 3},\mathbf{3})]$,
and to $[(\mathbf{8},\mathbf{1})\oplus(\mathbf{1},\mathbf{8})]$, and its ``mirror''
$[(\mathbf{1},\mathbf{8})\oplus(\mathbf{8},\mathbf{1})]$. Of course, one may
have other, ``exotic'' chiral multiplets that contain manifestly exotic flavor $SU_F(3)$
multiplets, but we exclude them {\it per fiat}, for lack of observed exotics.

Historically, after the observation, in Refs. \cite{Lee:1965,Dashen:1965im,Cabibbo:1966,Gerstein:1966bc,Oehme:1966},
that several crucial SU(6) algebra results follow from its (smaller) $SU(3) \otimes SU(3)$ sub-algebra, the notion
of $SU(3) \otimes SU(3)$ representation mixing was proposed as an explanation of the
nucleon's (isovector) axial coupling $g_A^{(1)}$. The physical nature of this
$SU(3) \otimes SU(3)$ sub-algebra was not immediately clear, however,
as two options (the conventional chiral charge algebra, and the so-called ``collinear'' algebra)
existed at the time.

Indeed, the ``collinear'' $SU(3) \otimes SU(3)$ algebra, which was generally assumed in the early work,
holds only in a particular (the so-called $p_z \to {\infty}$) frame of reference, which 
appears to be in conflict with the general principles of special relativity. 
Moreover, Adler and Weisberger had also derived their sum rule(s) with the help of the $p_{\infty}$ frame.
It was only after 
Weinberg's \cite{Weinberg:1972hw}
clarification of the Adler-Weisberger sum rule as consisting of two independent statements
({\it viz.} a) the model-independent Goldberger-Miyazawa-Oehme sum rule for the $\pi N$ scattering lengths;
and b) the chiral symmetry breaking-dependent predictions for the $\pi N$ scattering lengths)
that this matter was settled in favor of chiral symmetry,
and thus the way was paved for its later applications in QCD. Thus, only the chiral charge
symmetry option leads to Lorentz-invariant quark interaction theories, such as QCD.

\subsection{Chiral mixing}

In one of the earliest physical applications of the chiral configuration mixing idea,
Harari \cite{Harari:1966yq}, Bincer \cite{Bincer:1966pr},
Gerstein and Lee \cite{Gerstein:1966zz}, and Gatto et al. \cite{Gatto:1966pr}
used the mixing of three of the aforementioned six
chiral multiplets
to fit the nucleon's isovector axial coupling
constant $g_A^{(1)}$ value at 1.267, \cite{Yamanishi:2007zza} and thus 
explain its being different from unity, as was seemingly demanded
by the Gell-Mann--L\' evy model \cite{GellMann:1960np}. It turned out, however, that
this application was not selective at all: all mixing 
scenarios could reproduce this value, so long as the $[({\bf 6},{\bf 3})\oplus({\bf 3},{\bf 6})]$ multiplet
was involved:
a) Gatto {\it et al.} - Harari scenario \cite{Gatto:1966pr,Harari:1966yq}
\begin{eqnarray}
| N(8) \rangle &=& \sin \theta | ({\bf 6},{\bf 3}) \rangle +
\cos \theta (\cos \varphi | (\mathbf{3},\mathbf{\bar 3}) \rangle +
\sin \varphi | (\mathbf{\bar 3},\mathbf{3}) \rangle) \, ,
\label{e:Harari}
\end{eqnarray}
or
b) Gerstein-Lee scenario \cite{Gerstein:1966zz}
\begin{eqnarray}
| N(8) \rangle &=& \sin \theta | ({\bf 6},{\bf 3}) \rangle +
\cos \theta (\cos \varphi | (\mathbf{3},\mathbf{\bar 3}) \rangle +
\sin \varphi | (\mathbf{8},\mathbf{1}) \rangle) \, .
\label{e:Gerstein_Lee}
\end{eqnarray}
For other, more exotic scenarios, see Ref. \cite{Altarelli:1966ps}.
Simultaneously, or somewhat later, 
Refs. \cite{Gerstein:1966bc,Harari:1966jz,Gatto:1967pr,Weyers:1967pr,Love:1970vi},
used the same approach to saturate the electric dipole operator algebra and calculate
the nucleon's anomalous magnetic moments and charge radii. Moreover, 
other phenomenological applications of the current algebra e.g. to pion photoproduction
can be found in Ref. \cite{Gerstein:1966bc}. All of this was done in the framework of
collinear $SU(3) \otimes SU(3)$ algebra, but algebraically these results must be the same
as the chiral $SU(3) \otimes SU(3)$ algebra ones. The construction of corresponding
chiral multiplets in the $SU(3) \otimes SU(3)$ chiral charge algebra is not as straightforward
as in the collinear one, however (see our remarks about interpolators, below).

There is no guarantee that all six of the above chiral (not collinear) multiplets
are allowed by the Pauli principle in the ground state of the nucleon, as composed of three
Dirac quarks~\footnote{That aspect of the problem 
could be safely neglected 
in the collinear approach, which allows arbitrary values of the orbital angular
momentum $L$ and restricts only its z-projection $L_z$.}.
The (formal) tool for this kind of study was provided around 1980~\cite{Ioffe:1981kw,Chung:1981cc,Espriu:1983hu}, in the form of
the so-called nucleon-three-quark interpolating fields.

Studies, in Refs. \cite{Nagata:2007di,Nagata:2008zzc,Chen:2008qv,Chen:2012ex,Chen:2012vs}, of
local (S-wave, therefore ground state candidates), bi-local (P-wave and higher), and
tri-local (D-wave and higher) three-quark interpolators have shown that only
$[(\mathbf{\bar 3},\mathbf{3})\oplus(\mathbf{3},\mathbf{\bar 3})]$,
and $[(\mathbf{8},\mathbf{1})\oplus(\mathbf{1},\mathbf{8})]$ are allowed in the local limit
and that $[({\bf 6},{\bf 3})\oplus({\bf 3},{\bf 6})]$ appears as a spin 1/2 ``complement'' to
the local Rarita-Schwinger spin 3/2 interpolator.
Many other chiral multiplets appear in the non-local case, where the Pauli principle is less
restrictive.

\subsection{Isoscalar axial coupling}

The nucleon has also a flavor singlet axial coupling $g_A^{(0)}$,
that was measured/extracted from spin-polarized lepton-nucleon DIS data after 1988
as $g_A^{(0)}=$ 0.28 $\pm$ 0.16 \cite{Filippone:2001ux}, or the
more recent value of $0.33 \pm 0.03 \pm 0.05$ \cite{Bass:2007zzb},
which is in the non-relativistic quark model predicted to be unity.
Our studies of interpolating fields have shown that each
$SU_L(3) \times SU_R(3)$ multiplet carries definite $U_A(1)$ transformation
properties and the corresponding $U_A(1)$ charge, see Table \ref{tab:spin12}.
Then, the next basic question becomes if the same set of chiral mixing
angle(s) can simultaneously explain this anomalously low value.
The answer, which is in the positive~\cite{Dmitrasinovic:2011yf},
manifestly depends on the $U_A(1)$ chiral
transformation properties of the admixed nucleon fields, and leads to
the so-called Harari scenario that mixes $[({\bf 6},{\bf 3})\oplus({\bf 3},{\bf 6})]$,
with
$[(\mathbf{\bar 3},\mathbf{3})\oplus(\mathbf{3},\mathbf{\bar 3})]$, and its ``mirror''
$[(\mathbf{3},\mathbf{\bar 3})\oplus(\mathbf{\bar 3},\mathbf{3})]$ field.
No admixture of $[(\mathbf{8},\mathbf{1})\oplus(\mathbf{1},\mathbf{8})]$, or its
``mirror'' $[(\mathbf{1},\mathbf{8})\oplus(\mathbf{8},\mathbf{1})]$ is preferred.
This fact confirms the Gatto-Harari scenario, Eq. (\ref{e:Harari}), and eliminates the
Gerstein-Lee scenario, Eq. (\ref{e:Gerstein_Lee}), from contention.

Moreover, we note that the above outlined program of fitting the hadron/nucleon observables
in order to obtain chiral mixing angles is practically feasible only for the
ground state(s): e.g. there is no hope of ever (sufficiently accurately) measuring the
isovector axial coupling of the $\Delta$ resonance,
except, perhaps, on the lattice. The same comments hold for the negative parity, and all
of the higher-lying excited states. In this sense, the present scheme
is of limited scope, but its potential to explain and illustrate the (fairly
complex) QCD physics of baryons is undeniable.

The above no-$[(\mathbf{8},\mathbf{1})\oplus(\mathbf{1},\mathbf{8})]$, or
$[(\mathbf{1},\mathbf{8})\oplus(\mathbf{8},\mathbf{1})]$ 
selection rule is in striking agreement with the results of so-called
QCD sum rules and lattice QCD calculations~\cite{Ioffe:1981kw,Espriu:1983hu}
that indicate
only weak coupling of the physical nucleon ground state to the
$[(\mathbf{8},\mathbf{1})\oplus(\mathbf{1},\mathbf{8})]$,
and/or its ``mirror'' $[(\mathbf{1},\mathbf{8})\oplus(\mathbf{8},\mathbf{1})]$ multiplet component.
There is no dynamical, or symmetry-based explanation of this fact, as yet.

Specific dynamical models such as the Faddeev-Bethe-Salpeter-Schwinger-Dyson equation approach of
Ref. \cite{Eichmann:2009qa}, or the Faddeev-Salpeter equation approach of Ref. \cite{Loring:2001kv},
ought to yield specific predictions for these mixing angles/parameters, and perhaps
also to a dynamical explanation of empirical selection rules
such as the above one.

Irrespective of specific dynamical model calculations, there ought to exist an
effective chiral Lagrangian description of the corresponding hadron degrees of freedom.
The task of constructing them was long drawn out: a) the first $\left(1, \frac12 \right)-\left(0, \frac12 \right)$
chiral representation mixing Lagrangian with $SU_L(2) \otimes SU_R(2)$ chiral symmetry
was presented by Hara~\cite{Hara:1965}; b) the first ``naive''-``mirror''
$\left(\frac12 , 0 \right)-\left(0, \frac12 \right)$ chiral-mixing Lagrangian
with $SU_L(2) \otimes SU_R(2)$ chiral symmetry was presented by Lee~\cite{lee72}, and
further extended by a number of researchers \cite{Christos,DeTar:1988kn,Jido:1997yk,Jido:2001nt},
and most recently
by Nagata et al. \cite{Nagata:2008zzc,Nagata:2008xf,Dmitrasinovic:2009vy,Dmitrasinovic:2009vp};
c) the extension to $SU_L(3) \otimes SU_R(3)$ chiral symmetry
has been accomplished in
Refs. \cite{Christos2,Zheng:1991pk,Zheng:1992mn,Chen:2010ba,Chen:2011rh}
and will be briefly reviewed in Sect. \ref{sect:sigma model}.

\subsection{Consistency of chiral algebra}

In this paper we study the chiral properties of baryon states from the viewpoint of linear mixing of 
different chiral representations, which happens together with the chiral symmetry breaking. Before the chiral mixing,
it has been proved in Ref.~\cite{Chen:2009sf} that the axial charges of individual chiral multiplets
close the $SU_L(3) \otimes SU_R(3)$ chiral algebra (see also Appendix \ref{sect:SU(3)chiral algebra} of this paper). 
Unfortunately, we do not know how to use the chiral algebra to exactly describe the chiral symmetry breaking process, but we can 
still show that the axial charges obey the same $SU_L(3) \otimes SU_R(3)$ chiral algebra even after the chiral mixing has 
occurred, so the chiral mixing process can be well described by using the chiral algebra. After
we finetune the mixing parameters (angles) according to some experimental information, the obtained mixed nucleon state may be used
to study the physical nucleon state appeared after the chiral symmetry breaking, and at the same time be described by the chiral algebra.
Take the Gatto {\it et al.} - Harari scenario \cite{Gatto:1966pr,Harari:1966yq} as an example:
\begin{eqnarray}
\sin \theta | ({\bf 6},{\bf 3}) \rangle +
\cos \theta (\cos \varphi | (\mathbf{3},\mathbf{\bar 3}) \rangle +
\sin \varphi | (\mathbf{\bar 3},\mathbf{3}) \rangle) \xrightarrow{\rm chiral~mixing} | N(8) \rangle \xrightarrow{\rm chiral~symmetry~breaking} |N_{\rm phy} \rangle \, .
\end{eqnarray}
This will ensure that the chiral symmetry-breaking Dashen double commutator can be straightforwardly calculated.
We note that in general the physical chiral charges have some singular components, related to the chiral non-invariant QCD vacuum, but these are the chiral
charges after the chiral symmetry breaking, and we shall not study them in this paper.

\begin{table}
\caption{The Abelian and the non-Abelian axial charges (+ sign
indicates ``naive", - sign ``mirror" transformation properties)
and the non-Abelian chiral multiplets of $J^{P}=\frac12$, Lorentz
representation $(\frac{1}{2},0)$ nucleon fields. The field denoted
by $0$ belongs to the $(1,\frac12) \oplus (\frac12,1)$ chiral
multiplet and is the basic nucleon field that is mixed with
various $(\frac{1}{2},0)$ nucleon fields.} 
{\begin{tabular}{@{}lllllllll@{}} \toprule
case & field & $g_A^{(0)}$ & $g_A^{(1)}$ & $SU_L(2) \times SU_R(2)$ & F & D & $SU_L(3) \times SU_R(3)$ \\
\colrule I & $N_1 - N_2$ & $-1$ & $+1$ & $(\frac12,0) \oplus (0,\frac12)$ & $~~0$ & $+1$ & $(3,\overline{3}) \oplus (\overline{3}, 3)$ \\
II & $N_1 + N_2$ & $+3$ & $+1$ & $(\frac12,0) \oplus (0,\frac12)$ & $+1$ & $~~0$ & $(8,1) \oplus (1,8)$ \\
III & $N_1^{'} - N_2^{'}$ & $+1$ & $-1$ & $(0,\frac12) \oplus (\frac12,0)$ & $~~0$ & $-1$ & $(\overline{3},3) \oplus (3,\overline{3})$ \\
IV & $N_1^{'} + N_2^{'}$ & $-3$ & $-1$ & $(0,\frac12) \oplus (\frac12,0)$ & $-1$ & $~~0$ & $(1,8) \oplus (8,1)$ \\
\hline 0 & $\partial_{\mu}(N_3^{\mu} + \frac13 N_4^{\mu})$ & $+1$&
$+\frac53$ & $(1,\frac12) \oplus (\frac12,1)$ & $+\frac23$ & $+1$ & $(6,3) \oplus (3,6)$ \\
\botrule
\end{tabular}}
\label{tab:spin12}
\end{table}


A basic feature of the linear chiral realization is that the axial
couplings are determined by the chiral representations. In Ref.~\cite{Chen:2008qv}, we found that for the
nucleon octet,  the three-quark chiral representations
of $SU_L(3) \times SU_R(3)$,
$(\mathbf{8},\mathbf{1})\oplus(\mathbf{1},\mathbf{8})$,
$(\mathbf{3},\mathbf{\bar 3})\oplus(\mathbf{\bar 3},\mathbf{3})$ and
$(\mathbf{6},\mathbf{3}) \oplus (\mathbf{3},\mathbf{6})$ provide the
nucleon isovector axial coupling $g_A^{(3)} = 1$, 1 and $5/3$
respectively. Then in Ref.~\cite{Chen:2009sf}, we found that the mixing of chiral
$(\mathbf{6},\mathbf{3}) \oplus (\mathbf{3},\mathbf{6})$,
$(\mathbf{\bar 3},\mathbf{3})\oplus(\mathbf{3},\mathbf{\bar 3})$, and
$(\mathbf{3},\mathbf{\bar 3})\oplus(\mathbf{\bar 3},\mathbf{3})$ nucleons
leads to the observed axial couplings (the case III-I in Ref.~\cite{Chen:2009sf}):
\begin{align}
g_{A}^{(3)} &=
g_A^{(3)}(\mathbf{6},\mathbf{3}) \,
{\sin}^2 \theta + {\cos}^2\theta\,
\left(g_A^{(3)}(\mathbf{3},\mathbf{\bar 3})~ {\cos}^2 \varphi + g_A^{(3)}(\mathbf{\bar 3},\mathbf{3})~{\sin}^2\varphi \right) = 1.267 \, , \\
g_A^{(0)} &=
g_A^{(0)}(\mathbf{6},\mathbf{3})\, {\sin}^2 \theta
+ {\cos}^2 \theta\,
\left(g_A^{(0)}(\mathbf{3},\mathbf{\bar 3})~{\cos}^2 \varphi + g_A^{(0)}(\mathbf{\bar 3},\mathbf{3})~\,{\sin}^2 \varphi \right) = 0.33 \pm 0.08 \, ,
\end{align}
where we used
\begin{align}
\langle N | Q_{5}^{a} | N \rangle &= \langle N | Q_{5}^{a}(\mathbf{6},\mathbf{3})| N \rangle~\,{\sin}^2 \theta +
{\cos}^2\theta\,\left(\langle N | Q_{5}^{a}(\mathbf{3},\mathbf{\bar 3})| N \rangle
{\cos}^2 \varphi + \langle N | Q_{5}^{a}(\mathbf{\bar 3},\mathbf{3})| N \rangle{\sin}^2\varphi \right) \, .
\end{align}
Next we used Table \ref{tab:spin12} values of
$g_A^{(3)}(\mathbf{6},\mathbf{3}) = \frac{5}{3} = - g_A^{(3)}(\mathbf{3},\mathbf{6})$;
$g_A^{(3)}(\mathbf{3},\mathbf{\bar 3}) = 1 = - g_A^{(3)}(\mathbf{\bar 3},\mathbf{3})$
and $g_A^{(0)}(\mathbf{3},\mathbf{\bar 3}) = -1 = - g_A^{(0)}(\mathbf{\bar 3},\mathbf{3})$, to find
\begin{eqnarray}
g_{A}^{(3)} &=& \frac53~\sin^2\theta + \cos^2\theta ~ \cos 2 \varphi = 1.267 \, , \\
g_A^{(0)} &=& ~~{\sin}^2 \theta - {\cos}^2 \theta\, \cos 2 \varphi = 0.33 \pm 0.08 \,
, \label{e:axcoupl_observ}
\end{eqnarray}
whose solutions are
\begin{eqnarray}
\theta = 50.7^{\rm o} \pm 1.8^{\rm o} \, , \varphi = 66.1^{\rm o} \pm 2.9^{\rm o}  \, . \label{e:angles}
\end{eqnarray}
Of course, this mixing appears to affect the $SU_L(3) \otimes SU_R(3)$ chiral algebra, as well,
so we must first check that we did not spoil this algebra.
The main ``problematic'' part of the $SU_L(3) \otimes SU_R(3)$ chiral algebra
is the double-axial commutator
\begin{eqnarray}
\label{e:ax_ax}
[Q_5^a, Q_5^b]
&=&  i f^{abc} Q^c\, .
\end{eqnarray}
We shall check this commutation rule in the nucleon sub-space of the full Hilbert space:
\begin{eqnarray}
\label{e:commutator2}
\langle N |[Q_5^a, Q_5^b] | N \rangle
&=& \langle N |[Q_{5}^{a}(\mathbf{6},\mathbf{3}),Q_{5}^{b}(\mathbf{6},\mathbf{3})] | N \rangle~\sin^2\theta  \\
\nonumber
&+& \langle N |[Q_{5}^{a}(\mathbf{3},\mathbf{\bar 3}),Q_{5}^{b}(\mathbf{3},\mathbf{\bar 3})] | N \rangle
~\cos^2\theta ~{\cos}^2 \varphi\\
\nonumber
&+&
\langle N |[Q_{5}^{a}(\mathbf{\bar 3},\mathbf{3}), Q_{5}^{b}(\mathbf{\bar 3},\mathbf{3})] | N \rangle~\cos^2\theta ~{\sin}^2 \varphi \, .
\end{eqnarray}
Next, we may use the commutators $[Q_5^a, Q_5^b] = i f^{abc} Q^c$ for the $(\mathbf{3},\mathbf{\bar 3})$ and
the $(\mathbf{6},\mathbf{3})$ chiral multiplets
worked out in Ref. \cite{Chen:2009sf} and listed in Appendix \ref{sect:SU(3)chiral algebra}, which all
lead to the same SU(3) vector charges $Q^c$:
\begin{eqnarray}
&& [Q_5^a(\mathbf{3},\mathbf{\bar 3}), Q_5^b(\mathbf{3},\mathbf{\bar 3})] = i f^{abc} Q^c(\mathbf{3},\mathbf{\bar 3}) = i f^{abc} Q^c \, ,
\\ && [Q_5^a(\mathbf{6},\mathbf{3}), Q_5^b(\mathbf{6},\mathbf{3})] = i f^{abc} Q^c(\mathbf{6},\mathbf{3}) = i f^{abc} Q^c \, .
\end{eqnarray}
Thus we find
\begin{eqnarray}
\langle N |[Q_5^a, Q_5^b] | N \rangle
&=& i f^{abc} \langle N | Q^{c} | N \rangle~\left(\sin^2 \theta +
\cos^2\theta ~({\cos}^2 \varphi + \sin^2 \varphi)\right)
\nonumber \\
&=& i f^{abc} \langle N | Q^{c} | N \rangle \, ,
\end{eqnarray}
which confirms the chiral charge $SU_L(3) \otimes SU_R(3)$ algebra. This ensures that the
chiral symmetry-breaking Dashen double commutator can be safely/reliably calculated in the chiral
mixing approach.

\section{The linear sigma model for chiral mixing}
\label{sect:sigma model}

The next step is to try and reproduce this phenomenological mixing
starting from a model interaction, rather than {\it per fiat}. As
the first step in that direction we must look for a dynamical source
of chiral mixing. One, perhaps the simplest, such mechanism is the chirally
symmetric {\it non-derivative} one-$(\sigma,\pi)$-meson interaction
Lagrangian, which induces baryon masses via its $\sigma$-meson
coupling. For this reason we need to know the form of the most general
such Lagrangian(s); that problem was solved in Ref. \cite{Chen:2010ba}
for three flavors and in Refs. \cite{Dmitrasinovic:2009vp,Dmitrasinovic:2009vy}
for two flavors.

There is a significant difference between $N_f$=2 and $N_f$=3
chirally symmetric linear sigma models of chiral mixing, as
only in the latter case there are strongly restrictive selection rules.

For example, most $U_A(1)$ symmetry-breaking and $SU_L(3) \times SU_R(3)$ chiral
symmetry-conserving interactions are forbidden, see Tables \ref{tab:lags},\ref{tab:lagsN} taken from
Ref. \cite{Chen:2010ba}. In particular only one $SU_L(3) \times SU_R(3)$ symmetric,
but $U_A(1)$ symmetry-breaking interaction (the
$[(\mathbf{3},\mathbf{\bar 3})\oplus(\mathbf{\bar 3},\mathbf{3})]$-$[(\mathbf{8},\mathbf{1})\oplus(\mathbf{1},\mathbf{8})]$
and its Hermitian conjugate
$[(\mathbf{1},\mathbf{8})\oplus(\mathbf{8},\mathbf{1})]$[mir]-$[(\mathbf{\bar 3},\mathbf{3})\oplus(\mathbf{3},\mathbf{\bar 3}])$[mir])
is allowed. These results stand in marked contrast to the two-flavor
case~\cite{Dmitrasinovic:2009vp,Dmitrasinovic:2009vy}, where all of
the $SU_L(2) \times SU_R(2)$ symmetric interactions have both a
$U_A(1)$ symmetry-conserving and a $U_A(1)$ symmetry-breaking
version.
This is due to the fact
that in the $SU_L(2) \times SU_R(2)$ limit
both the $[(\mathbf{\bar 3},\mathbf{3})\oplus(\mathbf{3},\mathbf{\bar 3}])$
and the $[(\mathbf{8},\mathbf{1})\oplus(\mathbf{1},\mathbf{8})]$ multiplet
reduce to the same multiplet $(\half,0) \oplus (0,\half)$, albeit with different
$U_A(1)$ symmetry properties.

Although, the $SU_L(3) \times SU_R(3)$ symmetry is rather badly explicitly
broken, we may expect that in the corresponding $\pi - N$ sector, the
$SU_L(2) \times SU_R(2)$ symmetry may remain more-or-less conserved.
So, although we shall be primarily interested in the pion-nucleon
case, i.e., in $N_f$=2, we shall use the $N_f$=3 selection rules
for guidance.

\subsection{A brief Summary of $N_f$=3 Interactions}
\label{ssect:summary}

In this section, we introduce a shorthand notation
\begin{eqnarray}
{N}_{(8m)} &\sim& [(\mathbf{1},\mathbf{8})\oplus(\mathbf{8},\mathbf{1})]{\rm [mir]} \, ,
\nonumber \\
{N}_{(9m)} &\sim& [(\mathbf{\bar 3},\mathbf{3})\oplus(\mathbf{3},\mathbf{\bar 3})]{\rm [mir]} \, ,
\nonumber \\
{N}_{(18)} &\sim& (\mathbf{6},\mathbf{3}) \oplus (\mathbf{3},\mathbf{6}) \, ,
\nonumber \\
{N}_{(10m)} &\sim& [(\mathbf{1},\mathbf{10})\oplus(\mathbf{10},\mathbf{1})]{\rm [mir]} \, ,
\end{eqnarray}
and similar for their mirror and naive representations.
The scalar ($\sigma$) and pseudo-scalar ($\pi$) mesons are introduced and transformed
under the chiral transformations as
\begin{eqnarray}
M = \sigma + i \gamma_5 \pi \sim (\mathbf{3},\mathbf{\bar 3}), \; \; \;
M^\dagger = \sigma - i \gamma_5 \pi \sim (\mathbf{\bar 3},\mathbf{3}) \, .
\end{eqnarray}
Now the chiral structure of the Lagrangians for Yukawa type interactions is
\begin{eqnarray}
\bar N M N^\prime + \bar N^\prime M^\dagger N \sim
\bar N_L M N_R^\prime + \bar N_R^\prime M^\dagger N_L \, ,
\end{eqnarray}
where $N$ and $N^\prime$ may belong to different chiral representations.
Our task is to form chiral singlet combinations for these interactions.
For instance,
\begin{eqnarray}
\bar N_{(9m)} M N_{(18)} \sim
\overline{(\mathbf{\bar 3},\mathbf{3})} \otimes
(\mathbf{3},\mathbf{\bar 3}) \otimes
(\mathbf{3},\mathbf{6})
+
\overline{(\mathbf{3},\mathbf{\bar 3})} \otimes
(\mathbf{\bar 3},\mathbf{3}) \otimes
(\mathbf{6},\mathbf{3})
\label{eq_Lint_9m18}
\end{eqnarray}
can make the $SU_L(3) \times SU_R(3)$ chiral singlet
($\mathbf{3} \otimes  \mathbf{3} \otimes \mathbf{3} \to \mathbf{1},
\mathbf{\bar 3} \otimes  \mathbf{\bar 3} \otimes \mathbf{6} \to \mathbf{1})$.
This corresponds to the cell at the 2nd row and 3rd column of Table~\ref{tab:lags}.
Contrary,  a combination like
\begin{eqnarray}
\bar N_{(8m)} M N_{(8m)}
\sim
\overline{(\mathbf{1},\mathbf{8})}
\otimes (\mathbf{3},\mathbf{\bar 3}) \otimes
(\mathbf{8},\mathbf{1})
+
\overline{(\mathbf{8},\mathbf{1})}
\otimes (\mathbf{\bar 3},\mathbf{3}) \otimes
(\mathbf{1},\mathbf{8})
\end{eqnarray}
can not make the chiral invariant interaction as corresponding to the cell at the first row and column of
Table~\ref{tab:lags}.
We can also consider other possible combinations, all of which are listed in Table~\ref{tab:lags}.

\begin{table}[hbt]
\caption{Allowed chiral invariant interaction Lagrangian with one pseudoscalar meson field,
denoted  by either $M$ or $M^\dagger$ as corresponding to Eqs.~(\ref{eq:lag1}) and (\ref{eq:lag2}).
The symbol -- indicates that  chiral invariant construction is not allowed.
All cases are both $SU_L(3) \times SU_R(3)$ and $U_A(1)$ invariant except for
the last (third) group where $U_A(1)$ is broken.
}
\begin{center}
\label{tab:lags}
\begin{tabular}{l|c|c|c|c}
\hline \hline 
&
$(\mathbf{1},\mathbf{8})\oplus(\mathbf{8},\mathbf{1})$[mir] &
$(\mathbf{\bar 3},\mathbf{3})\oplus(\mathbf{3},\mathbf{\bar
3})$[mir] & $(\mathbf{6},\mathbf{3}) \oplus (\mathbf{3},\mathbf{6})$
& $(\mathbf{1},\mathbf{10})\oplus(\mathbf{10},\mathbf{1})$[mir]
\\
\hline
$\overline{(\mathbf{1},\mathbf{8}) \oplus (\mathbf{8},\mathbf{1})}$[mir] &
-- & $M^\dagger$ &  $M^\dagger$  & --
\\
$\overline{(\mathbf{\bar 3},\mathbf{3})\oplus(\mathbf{3},\mathbf{\bar 3})}$[mir]
&  $M^\dagger $&  $M$ &  $M$ & --
\\
$\overline{(\mathbf{6},\mathbf{3}) \oplus (\mathbf{3},\mathbf{6})}$
&  $M^\dagger$ &  $M$  &  $M$  &  $M^\dagger$
\\
$\overline{(\mathbf{1},\mathbf{{10}})\oplus(\mathbf{{10}},\mathbf{1})}$[mir]
& --& -- &  $M^\dagger$ & --
\\
\hline
\hline
& $(\mathbf{8},\mathbf{1})\oplus(\mathbf{1},\mathbf{8})$ &
$(\mathbf{3},\mathbf{\bar 3})\oplus(\mathbf{\bar 3},\mathbf{3})$
 & $(\mathbf{3},\mathbf{6})\oplus(\mathbf{6},\mathbf{3})$[mir] &
$(\mathbf{10},\mathbf{1}) \oplus (\mathbf{1},\mathbf{10})$
\\
\hline
$\overline{(\mathbf{8},\mathbf{1})\oplus(\mathbf{1},\mathbf{8})}$ &
-- & $M$ & $M$ & --
\\
$\overline{(\mathbf{\bar 3},\mathbf{3})\oplus(\mathbf{3},\mathbf{\bar 3})}$
   & $M$ & $M^\dagger$ & $M^\dagger$ & -- \\
$\overline{(\mathbf{\bar 3},\mathbf{\bar 6})\oplus(\mathbf{\bar 6},\mathbf{\bar 3})}$[mir]
  & $M$ & $M^\dagger$ & $M^\dagger$ & $M$
\\
$\overline{(\mathbf{\overline{10}},\mathbf{1}) \oplus (\mathbf{1},\mathbf{\overline{10}})}$ &
-- & -- &  $M$ & --
\\
\hline
\hline
&
$(\mathbf{8},\mathbf{1})\oplus(\mathbf{1},\mathbf{8})$ &
$(\mathbf{1},\mathbf{8})\oplus(\mathbf{8},\mathbf{1})$[mir] \\
\cline{1-3}
$\overline{(\mathbf{\bar 3},\mathbf{3})\oplus(\mathbf{3},\mathbf{\bar 3})}$
  & -- & $M^\dagger$; $U_A(1)$ broken
  \\
$\overline{(\mathbf{3},\mathbf{\bar 3})\oplus(\mathbf{\bar 3},\mathbf{3})}$[mir]
  &$M$;  $U_A(1)$ broken & --
\\ \hline \hline
\end{tabular}
\end{center}
\end{table}


The results are also expressed explicitly in the form of the Lagrangian which is given by
\begin{eqnarray}
\nonumber
\mathcal{L} &=& \left( \begin{array}{cccc}
\overline{N}_{(8m)} & \overline{N}_{(9m)} & \overline{N}_{(18)} &
\overline{N}_{(10m)}
\end{array} \right)
\Bigg(
M
\left (
\begin{array}{cccc}
{\bf 0}_{8\times8} & {\bf 0}_{8\times9} & {\bf 0}_{8\times18} &
{\bf 0}_{8\times10} \\
{\bf 0}_{9\times8} & g_{(9)}{\bf D}^a_{(9)} & g_{(9/18)}{\bf
T}^a_{(9/18)}& {\bf 0}_{9\times10} \\
{\bf 0}_{18\times8} & g^*_{(9/18)}{\bf T}^{\dagger a}_{(9/18)} &
g_{(18/18)}{\bf D}^a_{(18)}& {\bf 0}_{18\times10} \\
{\bf 0}_{10\times8} & {\bf 0}_{10\times9} & {\bf 0}_{10\times18} &
{\bf 0}_{10\times10} \end{array} \right) \\
&+&
M^\dagger
\left (
\begin{array}{cccc}
{\bf 0}_{8\times8} & g_{(8/9)}{\bf T}^a_{(8/9)} & g_{(8/18)}{\bf
T}^a_{(8/18)} & {\bf 0}_{8\times10} \\
g_{(8/9)}^*{\bf T}^{\dagger a}_{(8/9)} & {\bf 0}_{9\times9} &
{\bf 0}_{9\times18} & {\bf 0}_{9\times10} \\
g_{(8/18)}^*{\bf T}^{\dagger a}_{(8/18)} & {\bf 0}_{18\times9} &
{\bf 0}_{18\times18} & g_{(10/18)}^*{\bf T}^{\dagger a}_{(10/18)} \\
{\bf 0}_{10\times8} & {\bf 0}_{10\times9} & g_{(10/18)}{\bf
T}^a_{(10/18)} & {\bf 0}_{10\times10}
\end{array} \right)
\Bigg) \left ( \begin{array}{c} {N}_{(8m)} \\
{N}_{(9m)} \\ {N}_{(18)} \\ {N}_{(10m)}
\end{array} \right) \, .
\label{eq:lag1}
\end{eqnarray}
Here ${\bf 0}_{A \times B}$ is the null matrix of dimension ${A \times B}$, and
${\bf D}^a_{(A)} , {\bf T}^a_{(A/B)}$ are flavor transition matrix
of dimension as indicated by their subscripts, which are defined in Ref.~\cite{Chen:2009sf}.
Similarly the mirror counterparts are given as:
\begin{eqnarray}
\mathcal{L}_{(m)} \nonumber
&=& \left ( \begin{array}{cccc}
\overline{N}_{(8)} & \overline{N}_{(9)} & \overline{N}_{(18m)} &
\overline{N}_{(10)}
\end{array} \right) \Bigg(
M^\dagger
\left(\begin{array}{cccc}  {\bf 0}_{8\times8} & {\bf 0}_{8\times9} &
{\bf 0}_{8\times18} & {\bf 0}_{8\times10} \\
{\bf 0}_{9\times8} & g^\prime_{(9)}{\bf D}^a_{(9)} &
g^\prime_{(9/18)}{\bf T}^a_{(9/18)}& {\bf 0}_{9\times10} \\
{\bf 0}_{18\times8} & g^{\prime*}_{(9/18)}{\bf T}^{\dagger
a}_{(9/18)} & g^\prime_{(18/18)}{\bf D}^a_{(18)}& {\bf
0}_{18\times10} \\
{\bf 0}_{10\times8} & {\bf 0}_{10\times9} & {\bf 0}_{10\times18} &
{\bf 0}_{10\times10} \end{array} \right) \\
\label{eq:lag2}
&+&
M
\left(
\begin{array}{cccc}  {\bf 0}_{8\times8} & g^\prime_{(8/9)}{\bf T}^a_{(8/9)} &
g^\prime_{(8/18)}{\bf T}^a_{(8/18)} & {\bf 0}_{8\times10} \\
g_{(8/9)}^{\prime*}{\bf T}^{\dagger a}_{(8/9)} & {\bf
0}_{9\times9} & {\bf 0}_{9\times18} & {\bf 0}_{9\times10} \\
g_{(8/18)}^{\prime*}{\bf T}^{\dagger a}_{(8/18)} & {\bf
0}_{18\times9} & {\bf 0}_{18\times18} & g_{(10/18)}^{\prime*}{\bf
T}^{\dagger a}_{(10/18)} \\
{\bf 0}_{10\times8} & {\bf 0}_{10\times9} & g^\prime_{(10/18)}{\bf
T}^a_{(10/18)} & {\bf 0}_{10\times10} \end{array} \right ) \Bigg )
\left ( \begin{array}{c} {N}_{(8)} \\
{N}_{(9)} \\
{N}_{(18m)} \\ {N}_{(10)} \end{array} \right ) \, . \
\end{eqnarray}
Besides these, there is another single-term Lagrangian which
is also chiral invariant:
\begin{eqnarray}
\label{eq:lag3}
\mathcal{L}_{(B)} 
&=& g_{(B)}\overline{N}_{(8)}
M^\dagger
{\bf T}^a_{(B)} N_{(9m)} + h.c. \, ,
\end{eqnarray}
together with its mirror counterpart
\begin{eqnarray}
\label{eq:lag4}
\mathcal{L}_{(B m)} 
&=& g^\prime_{(B)}\overline{N}_{(8m)}
M
{\bf T}^{a}_{(B)} N_{(9)} + h.c. \, .
\end{eqnarray}
These corresponds to the third (bottom) group of Table~\ref{tab:lags}.

We note that the Lagrangians (\ref{eq:lag1}) and  (\ref{eq:lag2})
are also invariant under $U_A(1)$ chiral transformation, while
(\ref{eq:lag3}) and  (\ref{eq:lag4}) are not.
This is verified by counting the $U_A(1)$ charge $g_A^{(0)}$
in the interaction Lagrangian.
Recall that the meson fields $M$ and $M^\dagger$ carry
$g_A^{(0)} = -2$ and $+2$, respectively.
Therefore, for the interaction (\ref{eq_Lint_9m18}) as an example,
by using the result of Table~\ref{tab:spin12}
we have the net $U_A(1)$ charge as
\begin{eqnarray}
g_A^{(0)}
=
+1 -2 + 1 = 0\, ,
\end{eqnarray}
where we have used the fact that the $U_A(1)$ charge of the Dirac conjugate
is the same as the original one because of the interchange of the left and right components.
%

These results  stand in marked contrast to the two-flavor
case~\cite{Dmitrasinovic:2009vp,Dmitrasinovic:2009vy}.
Namely, for $SU_L(3) \times SU_R(3)$
chiral invariant Lagrangians with a certain representation structure as given
in Table~\ref{tab:lags} (or to one term in Eqs. (\ref{eq:lag1}) -- (\ref{eq:lag4}))
are either $U_A(1)$ symmetry-conserving or $U_A(1)$ symmetry-breaking.
In contrast, for $SU_L(2) \times SU_R(2)$
chiral invariant Lagrangians with the same representation structure
have both a $U_A(1)$ symmetry-conserving and a $U_A(1)$ symmetry-breaking
version.
This is due to the fact that the two $SU_L(3) \times SU_R(3)$ representations,
$(\mathbf{8},\mathbf{1}) \oplus (\mathbf{1},\mathbf{8})$ and
$(\mathbf{3},\mathbf{\bar 3}) \oplus (\mathbf{\bar 3},\mathbf{3})$
reduce to the same $(\half,0) \oplus (0,\half)$ representation of
$SU_L(2) \times SU_R(2)$.
Thus, generally speaking, the three-flavor chiral symmetry is more
restrictive than the two-flavor one.

Besides the interaction
Lagrangians (\ref{eq:lag1}-\ref{eq:lag4}),
the so-called ``naive''-``mirror'' mass terms are also chiral invariant:
\begin{eqnarray}
\mathcal{L}_{(mass)}
=
{m_{(8)}} \overline{N}_{(8m)} \gamma_5 N_{(8)} +
{m_{(9)}}\overline{N}_{(9m)} \gamma_5 N_{(9)}+
{m_{(18)}}\overline{N}_{(18m)} \gamma_5 N_{(18)}+
{m_{(10)}}\overline{N}_{(10m)} \gamma_5 N_{(10)} \, ,
\label{eq:Lmass}
\end{eqnarray}
where ${m_{(8)}},\cdots, {m_{(10)}}$  are the mass parameters.
The chiral structures of these terms are
summarized in Table~\ref{tab:lagsN}.

\begin{table}[hbt]
\caption{Allowed chiral invariant mass terms as denoted by 1, while
the symbol -- indicates that  chiral invariant construction is not allowed.
All cases are both $SU_L(3) \times SU_R(3)$ and $U_A(1)$ invariant.
}
\begin{center}\label{tab:lagsN}
\begin{tabular}{l|c|c|c|c}
\hline
\hline
($SU_A(3)$, $U_A(1)$) &
$(\mathbf{8},\mathbf{1})\oplus(\mathbf{1},\mathbf{8})$ &
$(\mathbf{3},\mathbf{\bar 3})\oplus(\mathbf{\bar 3},\mathbf{3})$ &
$(\mathbf{3},\mathbf{6})\oplus(\mathbf{6},\mathbf{3})$[mir] &
$(\mathbf{10},\mathbf{1})\oplus(\mathbf{1},\mathbf{10})$
\\
\hline
$\overline{(\mathbf{1},\mathbf{8})\oplus(\mathbf{8},\mathbf{1})}$[mir] &
  1 &-- & -- & --
\\
$\overline{(\mathbf{3},\mathbf{\bar 3})\oplus(\mathbf{\bar 3},\mathbf{3})}$[mir] &
  -- & 1 & -- & --
\\
$\overline{(\mathbf{\bar 6},\mathbf{\bar 3}) \oplus (\mathbf{\bar 3},\mathbf{\bar 6})}$ &
  -- & -- & 1 & --
\\
$\overline{(\mathbf{1},\mathbf{\overline{10}})\oplus(\mathbf{\overline{10}},\mathbf{1})}$[mir]&
  -- & -- & -- & 1
\\
\hline
\hline
\end{tabular}
\end{center}
\end{table}

\subsection{Baryon Masses in the chiral limit}
\label{ssect:masses}

Chiral symmetry is spontaneously broken through the
``condensation'' of the sigma field $\sigma \rightarrow
\sigma_0=\langle \sigma \rangle_0 = f_\pi$, which leads to the
dynamical generation of baryon masses, as can be seen from the
linearized chiral invariant interaction Lagrangians
Eqs.~(\ref{eq:lag1}), (\ref{eq:lag2}), (\ref{eq:lag3}) and (\ref{eq:lag4}).

In this section, we study the masses of the octet baryons. There are
altogether six types of octet baryon fields: $N_+$ ($N_{(8)}$),
$N_-$ (contained in $N_{(9)}$) and $N_\mu$ (contained in
$N_{(18)}$), as well as their mirror fields $N_+^{\prime}$
($N_{(8m)}$), $N_-^{\prime}$ (contained in $N_{(9m)}$),
$N^\prime_\mu$ (contained in $N_{(18m)}$). The nucleon mass matrix
is already in a simple block-diagonal form when the nucleon fields
form the following mass matrix:
\begin{eqnarray}
M &=& {1\over\sqrt{6}}\bar N \left ( \begin{array}{ccc|ccc} 0 &
f_\pi g_{(8/9)} & f_\pi g_{(8/18)} & m_{(8)} \gamma_5 & f_\pi g_{B} & 0
\\
f_\pi g^*_{(8/9)} & f_\pi g_{(9/9)} & f_\pi g_{(9/18)} & f_\pi g^*_{B} & m_{(9)} \gamma_5 & 0
\\
f_\pi g^*_{(8/18)} & f_\pi g^*_{(9/18)}& f_\pi g_{(18/18)} & 0 & 0 & m_{(18)} \gamma_5
\\ \hline
m_{(8)} \gamma_5 & f_\pi g^\prime_{B} & 0 & 0 & f_\pi g^\prime_{(8/9)} &
g^\prime_{(8/18)}
\\
f_\pi g^{\prime*}_{B} & m_{(9)} \gamma_5 & 0 & f_\pi g^{\prime*}_{(8/9)} &
f_\pi g^\prime_{(9/9)} & f_\pi g^\prime_{(9/18)}
\\
0 & 0 & m_{(18)} \gamma_5 & f_\pi g^{\prime*}_{(8/18)} &
f_\pi g^{\prime*}_{(9/18)}& f_\pi g^\prime_{(18/18)}
\end{array} \right ) N \, ,
\end{eqnarray}
where
\begin{eqnarray}
N = ( N_+^{\prime}, N_-^{\prime}, N_\mu, N_+, N_-, N_\mu^\prime )^T
\, .
\end{eqnarray}
Since there are three nucleon fields as well as their mirror fields,
there can be a nonzero phase angle. However, for simplicity, we
assume all the axial couplings are real.

\subsection{Masses due to $[({\bf 6},{\bf 3})\oplus({\bf
3},{\bf 6})]$--$[(\mathbf{\bar 3},\mathbf{3})\oplus(\mathbf{
3},\mathbf{\bar 3})]$--$[(\mathbf{3},\mathbf{\bar 3})\oplus(\mathbf{
\bar 3},\mathbf{3})]$ mixing} \label{ssect:(6,3)(3,3)masses}

As shown in Sec.~\ref{sect:Phenomenology}, the mixing of chiral
$(\mathbf{6},\mathbf{3}) \oplus (\mathbf{3},\mathbf{6})$,
$(\mathbf{\bar 3},\mathbf{3})\oplus(\mathbf{3},\mathbf{\bar 3})$, and
$(\mathbf{3},\mathbf{\bar 3})\oplus(\mathbf{\bar 3},\mathbf{3})$ nucleons
leads to the observed axial couplings (the case III-I in Ref.~\cite{Chen:2009sf}).
Accordingly, we investigate the following three nucleon chiral multiplets
\begin{eqnarray}
(B_2,\Delta) &\in& ({\bf 6},{\bf 3})\oplus({\bf 3},{\bf 6}) \, ,
\nonumber \\
\label{eq:3multiplets}
(B_1,\Lambda_1) &\in& (\mathbf{\bar 3},\mathbf{3})\oplus(\mathbf{ 3},\mathbf{\bar 3}) [{\rm mir}] \, , \\
\nonumber
(B_3,\Lambda_2) &\in& (\mathbf{3},\mathbf{\bar 3})\oplus(\mathbf{\bar 3},\mathbf{ 3}) \, ,
\end{eqnarray}
and one meson multiplet
\begin{eqnarray}
\nonumber
(\sigma,\pi) &\in& (\mathbf{3},\mathbf{\bar 3})\oplus(\mathbf{\bar 3},\mathbf{3}) \, .
\end{eqnarray}
Here all baryons have spin 1/2, while the isospin of $B_1$ and $B_2$
is 1/2 and that of $\Delta$ is 3/2. The $\Delta$ field is then
represented by an isovector, Dirac-spinor field $\Delta^i$,
($i=1,2,3$), which should not be confused with the spin-$\frac32$ $\Delta(1232)$ resonance.

In writing down the Lagrangians Eqs.~(\ref{eq:lag1}), 
we have implicitly
assumed that the parities of $B_1$, $B_2$, $\Lambda$ and $\Delta$
are the same. In principle, they are arbitrary, except for the
ground state nucleon, which must be even. For instance, if $B_2$
has odd parity, the first term in the interaction Lagrangian
Eq.~(\ref{eq:lag1}) must include another $\gamma_5$
matrix~\cite{Jido:2001nt}.


Having established the mixing interactions
as well as the diagonal terms in Ref. \cite{Chen:2010ba},
we calculated the masses of the baryon states,
as functions of the pion decay constant $f_\pi$
and
the coupling constants $g_1 \sim g_{(9/9)}$, $g_2 \sim g_{(18/18)}$ and
$g_3 \sim g_{(9/18)}$:
\begin{eqnarray}
\label{def:33}
\nonumber \mathcal{L}_{(9)} &=& - g_1 f_\pi \Big ( \bar B_1 B_1 - 2
\bar \Lambda  \Lambda \Big ) + \cdots \, ,
\\ \mathcal{L}_{(18)} &=& - g_2 f_\pi \Big ( \bar B_2  B_2 -
2 \bar \Delta^i  \Delta^i \Big ) + \cdots \, ,
\\ \nonumber \mathcal{L}_{(9/18)} &=& - g_3 f_\pi \Big ( \bar B_1  B_2 \Big ) + \cdots \, , \\
\label{def:33_2} \nonumber \mathcal{L}^\prime_{(9)} &=& - g_4 f_\pi \Big (
\bar B_3  B_3 - 2 \bar \Lambda_1  \Lambda_1 \Big ) + \cdots \, ,\\
\mathcal{L}_{(9/9)} &=& - g_5 f_\pi \bar B_1 B_3 - g_5 f_\pi \bar
\Lambda_1 \Lambda_2 + \cdots \, .
\end{eqnarray}
We note that $B_1$ and $B_3$
couple with each other through the naive combinations:
${m_{(9)}}\overline{N}_{(9m)} \gamma_5 N_{(9)}$.
Altogether we have
\begin{eqnarray}
\mathcal{L} &=& - f_\pi (\bar B_1, \bar B_3 , \bar B_2) \left(
\begin{array}{ccc} g_1 & g_5 & g_3
\\ g_5 & g_4 & 0
\\ g_3 & 0 & g_2
\end{array} \right ) \left(
\begin{array}{c} B_1
\\ B_3
\\ B_2
\end{array} \right ) - f_\pi (\bar \Lambda_1 , \bar \Lambda_2) \left(
\begin{array}{cc} - 2 g_1 & g_5
\\ g_5 & - 2 g_4
\end{array} \right ) \left(
\begin{array}{c} \Lambda_1
\\ \Lambda_2
\end{array} \right ) + 2 g_2 f_\pi \bar \Delta^i \Delta^i \, . \nonumber
\\
\end{eqnarray}

Let us now diagonalize the mass matrix and express the mixing angle in terms
of diagonalized masses.
We use the three nucleon candidates
$N(940)$, $N(1440)$ and $N^*(1535)$ as well as the two mixing
angles $\theta = 50.7^{\rm o}$ and $\varphi = 66.1^{\rm o}$, and finally find that
there are two possibilities as shown in
Table~\ref{tab:prediction4}~\cite{Chen:2010ba}.
The odd-parity $\Delta$ option appears as the better one.
Now, the first flavor-singlet $\Lambda$ lies at 1370 MeV,
substantially closer to 1405 MeV.
flavor-singlet $\Lambda$ lies at 1850 MeV, very close to the
(three star PDG~\cite{Amsler:2008zzb}) $P_{01}(1810)$
resonance. This is our best candidate in the $[({\bf 6},{\bf
3})\oplus({\bf 3},{\bf 6})]$--$[(\mathbf{\bar
3},\mathbf{3})\oplus(\mathbf{ 3},\mathbf{\bar
3})]$--$[(\mathbf{3},\mathbf{\bar 3})\oplus(\mathbf{ \bar
3},\mathbf{3})]$ mixing scenario.

\begin{table}[tbh]
\begin{center}
\caption{The values of the $\Delta$ and $\Lambda$ baryon masses
predicted from the isovector axial coupling $g_{A~\rm mix.}^{(1)}
= g_{A~ \rm expt.}^{(1)} = 1.267$ and $g_{A~\rm mix.}^{(0)} = 0.33
\pm 0.08$ due to $[({\bf 6},{\bf 3})\oplus({\bf 3},{\bf 6})]$ --
$[(\mathbf{\bar 3},\mathbf{3})\oplus(\mathbf{ 3},\mathbf{\bar
3})]$ -- $[(\mathbf{3},\mathbf{\bar 3})\oplus(\mathbf{ \bar
3},\mathbf{3})]$ mixing .}
\begin{tabular}{c|ccccc|ccc}
\hline \hline No. & $g_1$ & $g_2$ & $g_3$ & $g_4$ & $g_5$ &
$\Lambda_1^P$ (MeV) & $\Lambda_2^P$ (MeV) & $\Delta^P$ (MeV) \\
\hline 1 & $-4.7$ & 8.4 & $-3.4$ & 2.9 & 9.8 & $1370^-$ & $1850^+$ &
$2170^-$
\\ 2 & $-7.2$ & 4.6 & 7.9 & 9.1 & $-4.2$ & $1940^+$ & $2430^-$ & $1200^-$
\\ \hline \hline
\end{tabular}
\label{tab:prediction4}
\end{center}
\end{table}

A comment about the comparatively high value of the $\Delta$ mass
is in order: In the mid-1960-s Hara \cite{Hara:1965}
noticed that the chiral transformation rules for a $(1,\half)$
multiplet impose a strict and seemingly improbable mass relation
among its two members: $m_{\Delta} = 2 m_{N}$. The mixing with the
$(\half,0)$ multiplet only makes things worse,
i.e., it makes the $\Delta$ even heavier. For this reason, the
lowest-lying spin-1/2 $\Delta$ resonance cannot be a chiral partner
of the lowest-lying nucleon $N$(940), whereas, $\Delta(2150)$ seems
to be a viable candidate for the $N$(940)'s chiral partner. Of course,
$\Delta(2150)$ may contain components of (i.e. mix with) other
high-lying resonances that do not significantly mix with $N$(940).

\subsection{Chiral symmetry breaking}
\label{ssect:current_masses}

\subsubsection{Chiral symmetry breaking: bare quark masses}
\label{ssect:quark_current_masses}

In QCD one expects ${\cal H}_{\rm \chi SB}$ to be determined solely by the current quark
masses $m_u^0, m_d^0, m_s^0$ (modulo EM effects), i.e.,  ${\cal H}_{\chi SB} = {\cal H}^{\rm q}_{\chi SB}$:
\begin{eqnarray}
\label{e:HsbChengLi}
{\cal H}^{\rm q}_{\chi SB}
&=& m_u^0 \bar u u + m_d^0 \bar d d + m_s^0 \bar s s \\
\nonumber 
&=&
\sum_{i=u,d,s} {\bar q}_i m_{q_i}^{0} q_i \\ \nonumber
&=& \sum_{a=0,3,8} m_{a}^{0} ({\bar q} {\bf \lambda}^{a} q)\, ,
\end{eqnarray}
where
\begin{eqnarray}
\nonumber 
m_{0}  &=& \frac{m_d+m_s+m_u}{\sqrt{6}} \, , \\
\nonumber
m_{3}  &=& \frac{1}{2} \left(m_u-m_d\right) \, , \\
\nonumber
m_{8}  &=& \frac{m_d-2 m_s+m_u}{2 \sqrt{3}} \, .
\nonumber
\end{eqnarray}

\subsubsection{Chiral symmetry breaking: bare baryon masses}
\label{ssect:baryon_current_masses}

We introduce, following Refs.~\cite{Dmitrasinovic:1999mf,Weinberg1996},
an explicit $\chi$SB ``bare'' nucleon mass and the corresponding $\chi$SB
Hamiltonian density. 
\begin{eqnarray}
\nonumber 
{\cal H}^{\rm N(full)}_{\chi SB}  &=&
\sum_{i,j=1}^{3} {\bar N}_i M_{N_{ij}}^{0} N_j
+\bar \Delta_{(1,\frac12)} M_{\Delta(1,\frac12)}^0 \Delta_{(1,\frac12)} \, ,
\end{eqnarray}
where $i$ stands for the three chiral multiplets
$(1,\frac12)$, $(\frac12,0)$ and $(0,\frac12)$. 
The off-diagonal components ($i \neq j$) needs not be zero, but we find that only the
diagonal components ($i = j$) contribute to the $\Sigma$ term, which will be calculated
in the next section, Sec.~\ref{sect:pi_Sigma_term}. Hence, we only pay attention to
the diagonal Hamiltonian
\begin{eqnarray}
\nonumber 
{\cal H}^{\rm N}_{\chi SB}  &=&
\sum_{i=1}^{3} {\bar N}_i M_{N_i}^{0} N_i
+\bar \Delta_{(1,\frac12)} M_{\Delta(1,\frac12)}^0 \Delta_{(1,\frac12)} \, .
\end{eqnarray}
{\it A priori}, we do not know the values
of the ``current'' nucleon masses, except for a lower limit - they cannot be smaller
than three isospin-averaged current quark masses:
$M_{N_i}^{0} \geq 3 {\bar m}_{q}^{0}$ = ${3 \over 2}\left(m_{u}^{0} + m_{d}^{0}\right) \simeq$
23 MeV~\cite{Caso:1998tx}, or 14 MeV~\cite{Beringer:1900zz}.

To see how this bound comes about, note that
the isospin-averaged ``bare'' nucleon mass term,
\begin{eqnarray}
\mathcal{H}_{\rm \chi SB}(0) = M_{N}^0 \bar N N \, ,
\end{eqnarray}
where
${\bar M}_{N}^{0} = {1 \over 2} \left(M_{p}^{0} + M_{n}^{0}\right)$,
can be readily expressed in terms of the current quark mass term, Eq. (\ref{e:HsbChengLi}),
with $M_{N}^{0} = 3 {\bar m}_{q}^{0}$ = ${3 \over 2}
\left(m_{u}^{0} + m_{d}^{0}\right)$.

It seems clear that the same ``current'' (or bare) nucleon mass $M_{N}^{0}$
ought to hold for any of the three chiral multiplets $(1,\frac12)$, $(\frac12,0)$
and $(0,\frac12)$, so long as they all correspond to three-quark interpolating fields.
Of course, the same chiral multiplets may arise as five-quark interpolators, in which
case their bare mass ought to be ${5 \over 2} \left(m_{u}^{0} + m_{d}^{0}\right)$, i.e.,
larger than the above value ${3 \over 2} \left(m_{u}^{0} + m_{d}^{0}\right)$. That explains
the inequality in $M_{N_i}^{0} \geq 3 {\bar m}_{q}^{0} = {3 \over 2}\left(m_{u}^{0} + m_{d}^{0}\right)$.

For simplicity's sake, we shall assume, as a first approximation, that all
three chiral components have the same ``current'' nucleon mass
$M_{N}^0 = M_{(6,3)}^0 = m_N^{(1,\frac12)} = m_{\Delta}^{(1,\frac12)}
= M_{(3,\bar3)}^0 = m_N^{(\frac12,0)} = M_{(\bar3,3)}^0 = m_{N}^{(0,\frac12)} =
{3 \over 2}\left(m_{u}^{0} + m_{d}^{0}\right)$.
In principle, the nucleon bare mass value may differ from one chiral multiplet to another, albeit not
by much, e.g. in the three-flavor case it may contain different $F$ and $D$ components,
due to different $F$ and $D$ structures of the chiral multiplets, see below.
This difference may be important in the three-flavor extension(s) of the model,
but not in the two-flavor case.

The model is easily extended to broken SU(3) symmetry case:
the explicit $\chi$SB ``bare'' nucleon mass and the corresponding diagonal $\chi$SB
Hamiltonian density are
\begin{eqnarray}
\label{e:Nucleon_mass}
{\cal H}^{\rm N}_{\chi SB}  &=&
\sum_{i=1}^{3} {\bar B}_i M_{B_i}^{0} B_i
+\bar \Delta_{(6,3)} M_{\Delta(6,3)}^0 \Delta_{(6,3)}  \, ,
\end{eqnarray}
where $i$ stands for the three chiral multiplets $({\bf 6},{\bf 3})$,
$(\mathbf{\bar 3},\mathbf{3})$ and $(\mathbf{3},\mathbf{\bar 3})$, and
the nucleon-octet mass matrix $M_{B_i}^{0}$ in the ``physical'' basis
reads
\begin{eqnarray}
\label{e:Nucleon_mass2}
M_{B_i}^{0} &=&
m_3 \left(d_3 + f_3 \right) - \frac{3}{2 \sqrt{3}} d_8
(M_{\Lambda} - M_{\Sigma}) + 2 f_8 m_8 + \sqrt{6} m_0 U \, , \\
\label{e:Nucleon_mass3}
M_{B_i}^{0} &=& {\rm diagonal}
\left(
\begin{array}{c}
 \frac{3 (M_{\Lambda} - M_{\Sigma})}{4} + m_d + 2 m_u, \\
 \frac{3 (M_{\Lambda} - M_{\Sigma})}{4}+2 m_d+m_u, \\
 \frac{1}{2} \left(m_d + 2 m_s + 3 \left(-M_{\Lambda} + M_{\Sigma} + m_u \right)\right) \\
-\frac{3 M_{\Lambda} }{2}+ \frac{3 M_{\Sigma}}{2} + m_d + m_s + m_u, \\
\frac{1}{2} \left(-3 M_{\Lambda} + 3 M_{\Sigma} + 3 m_d + 2 m_s + m_u\right), \\
\frac{1}{4} \left(3 M_{\Lambda} - 3 M_{\Sigma} + 2 m_d + 8 m_s + 2 m_u\right), \\
\frac{1}{4} \left(3 M_{\Lambda} - 3 M_{\Sigma} + 2 m_d + 8 m_s + 2 m_u\right), \\
\frac{3 (M_{\Lambda} - M_{\Sigma})}{2}  +m_d+m_s+m_u 
\end{array}
\right) + {\rm off}{\rm -}{\rm diagonal} 
\, .
\end{eqnarray}
The off-diagonal term in Eq. (\ref{e:Nucleon_mass3}), given by
\begin{eqnarray}
\nonumber 
{\rm off}{\rm -}{\rm diagonal} 
&=& \left(\frac{m_u-m_d}{2 \sqrt{3}} \right)
\left(
\begin{array}{cccccccc}
 0 & 0 & 0 & 0 & 0 & 0 & 0 & 0 \\
 0 & 0 & 0 & 0 & 0 & 0 & 0 & 0 \\
 0 & 0 & 0 & 0 & 0 & 0 & 0 & 0 \\
 0 & 0 & 0 & 0 & 0 & 0 & 0 & 1 \\
 0 & 0 & 0 & 0 & 0 & 0 & 0 & 0 \\
 0 & 0 & 0 & 0 & 0 & 0 & 0 & 0 \\
 0 & 0 & 0 & 0 & 0 & 0 & 0 & 0 \\
 0 & 0 & 0 & 1 & 0 & 0 & 0 & 0 
\end{array}
\right)
\, ,
\end{eqnarray}
determines the $\Lambda - \Sigma_0$ mixing and mass splitting.
As we shall not concern ourselves with 
hyperons in this paper, this term is of no interest here.

\section{The pion-nucleon $\Sigma_{\pi N}$ term}
\label{sect:pi_Sigma_term}

\subsection{Introduction: $\Sigma_{\pi N}$ at quark level}
\label{ss:piN_Sigma_Introduction}

The $\Sigma$ operator, defined as the double commutator
\begin{eqnarray}
\label{e:Sigma_def}
\Sigma &=&
\frac13 \delta^{ab} [ Q_5^a, [ Q_5^b, \mathcal{H}_{\rm \chi SB}(0) ] ]  \, ,
\end{eqnarray}
was introduced by Dashen as a measure of explicit chiral $SU_L(2) \times SU_R(2)$
symmetry breaking~\cite{dash69,reya74,Cheng:1970mx,Li:1971vr}.
It is sensitive to the 
flavor indices
of the axial charges $Q^{a}_{5}$ and the form of the $SU_L(3) \times SU_R(3)$
chiral symmetry breaking Hamiltonian density ${\cal H}_{\chi SB}$:
Other choices of summed over indices $a,b$ probe different parts of
symmetry breaking Hamiltonian.
Its nucleon matrix element is the pion-nucleon $\Sigma_{\pi N}$ term
\begin{eqnarray}
\label{e:Sigma_piN_def}
\Sigma_{\pi N} &=&
\frac13 \delta^{ab} \langle N | [ Q_5^a, [ Q_5^b, \mathcal{H}_{\rm \chi SB}(0) ] ]
| N \rangle \, ,
\end{eqnarray}
which is of importance for the determination of the flavour content, in particular of the
$s \bar s$ content of the nucleon~\cite{Cheng:1970mx,Cheng:1975wm,Donoghue:1985bu,kj87}.
In QCD one expects ${\cal H}_{\chi SB}$ to be determined solely by the current quark
masses $m_u^0, m_d^0, m_s^0$ (modulo EM effects) Eq. (\ref{e:HsbChengLi}).
Then, the axial charges $Q_5^a$ are then also constructed from the quark fields:
\begin{eqnarray}
Q_5^a = \int d{\bf x} q^{\dagger}(x) \gamma_5 {1\over 2}\lambda^a q(x) \, ,
\end{eqnarray}
which leads, after some basic algebraic manipulations, to
\begin{eqnarray}
\Sigma_{\pi N} = {m_u^0 + m_d^0 \over 2} \langle N | \bar u u + \bar d d | N \rangle
+ m_s^0 \langle N | \bar s s | N \rangle \, ,
\end{eqnarray}
and thus the value of the $\pi N$ $\Sigma_{N}$-term that is
given by the sum of the current quark masses in the nucleon.
(This is reflected in a nonzero ``bare'', or ``current'' nucleon mass
on the hadronic level.)
Assuming the nucleon contains no, or little strange quark component, i.e.,
$\langle N | \bar s s | N \rangle \sim 0$, one has
\begin{eqnarray}
\Sigma_{\pi N} = {m_u^0 + m_d^0 \over 2} \langle N | \bar u u + \bar d d | N \rangle \, .
\end{eqnarray}
The matrix element $\langle N | \bar u u + \bar d d | N \rangle$ counts the number
of $u$ and $d$ quarks and/or antiquarks in the nucleon, so that
$\Sigma_{\pi N} \simeq {3 \over 2}(m_u^0 + m_d^0) = 3 \hat m^0 \simeq 23$ MeV (with
the current quark mass estimates that were valid at the time in PDG1998~\cite{Caso:1998tx}; these have dropped in the
meantime significantly down to roughly $3 \hat m^0 \simeq 14$ MeV in PDG2012~\cite{Beringer:1900zz}).
At the same time the nucleon mass shift due to the $SU(3)$-breaking Hamiltonian
was evaluated at the baryonic level, assuming the contribution of strange quark
to be zero, as
\begin{eqnarray}
\Sigma_{\pi N} = {3 \hat m^0 \over m_s^0 - \hat m^0} (M_\Xi - M_\Lambda)
\simeq 26 ~{\rm MeV} \, ,
\end{eqnarray}
where $M_\Xi, M_\Lambda$ are the hyperon ground state masses.
As these two essentially independent estimates yielded basically one and the same
number, any
deviation of $\Sigma_{\pi N}$ from the value of 25 MeV seemed to indicate some
$\bar s s$ content in the nucleon (this agreement between these two methods has disappeared
with time, however: with the PDG2012~\cite{Beringer:1900zz} values one finds
$\Sigma_{\pi N} = {3 \hat m^0 \over m_s^0 - \hat m^0} (M_\Xi - M_\Lambda) \simeq 22 ~{\rm MeV}$ vs.
$\Sigma_{\pi N} \simeq {3 \over 2}(m_u^0 + m_d^0) \simeq 13$ MeV).
But, all estimates of $\Sigma_{\pi N}$ from the $\pi N$ scattering data yielded substantially
larger values, ranging from 55 MeV to 80 MeV~\cite{Borasoy:1996bx,Pavan:2001wz,Hite:2005tg,Alarcon:2011zs}.
Consequently, the importance of the $\Sigma_{\pi N}$ term cannot be exaggerated for the $s \bar s$
content of the nucleon. These arguments go back to 1976~\cite{Cheng:1975wm}, and have, by now, found their way into
textbooks on particle physics~\cite{Cheng:1985bj,Donoghue:1992dd}.

In the meantime there has been a large number of attempts at a theoretical explanation,
most of which rely on the enlarged $s \bar s$ content of the nucleon.
More recently a number of lattice calculations with (almost physical) pions have also
reached an enlarged value of $\Sigma_{\pi N}$~\cite{Ohki:2008ff,Babich:2010at,Bali:2011ks,Dinter:2012tt}.
But, there have also been many experimental searches for the $s \bar s$ contributions to
the nucleon observables, none of which produced a significant result (meaning larger than
1\% of the $u \bar u$ and $d \bar d$ contributions; otherwise they are compatible with
isospin violating corrections) thus making $s \bar s$ effectively negligible~\cite{Acha:2006my,Baunack:2009gy,Androic:2009aa}.
Thus the enigma deepens: how is it possible to have such a large $\Sigma$ term
without $s \bar s$ content?

\subsection{$\Sigma_{\pi N}$ at the baryon level}
\label{ss:piN_Sigma_mass}

The results obtained at the quark level are not always the same as those obtained
at the hadronic level, however.
The purpose of this study is to lay bare the dependence of the $\Sigma_ {\pi N}$ term on
the mixing of chiral multiplets, i.e., on the isovector axial coupling
$g_A^{(3)}$, and the flavor-singlet axial coupling $g_A^{(0)}$.

The axial current coupling constants of the baryon flavor octet
are well known~\cite{Yamanishi:2007zza}. The zeroth
(time-like) components of these axial currents are generators of
the $SU_L(3) \times SU_R(3)$ chiral symmetry of QCD.
The general flavor $SU_F(3)$ symmetric form of the nucleon
axial current contains two free parameters, called the $F$ and $D$
couplings, that are empirically determined as $F$=$0.459 \pm
0.008$ and $D$=$0.798 \pm 0.008$~\cite{Yamanishi:2007zza}.
The nucleon also has a flavor singlet axial coupling $g_A^{(0)}$,
that has been estimated from spin-polarized lepton-nucleon DIS
data as $g_A^{(0)}=$ 0.28 $\pm$ 0.16~\cite{Filippone:2001ux}, or
more recently as $0.33 \pm 0.03 \pm 0.05$~\cite{Bass:2007zzb,Vogelsang:2007zza}, subject
to certain assumptions about hyperon decays (the axial F and D values).

In the chiral mixing approach the value $g_A \ne 1$ is achieved naturally
by way of mixing different chiral multiplets, without derivative couplings.
We shall display the $\Sigma_{\pi N}$ term's dependence on $g_A^{(3)}$ and show that a large
value of $\Sigma_{\pi N}$ is easily obtained even with the present day
(significantly smaller) current quark masses and with a vanishing $s\bar{s}$
component of the nucleon.

First, we discuss the nucleon $\Sigma_{\pi N}$ term as obtained from the $\Sigma$
double commutator. We adopt different chiral symmetry breaking
[$\chi$SB] terms, 
in accord with Refs. \cite{camp79,bc79}.
Then we evaluate the $\Sigma_{\pi N}$ term in this approach.


\subsubsection{Chiral $SU_L(3) \times SU_R(3)$ Commutators}

We note that the matrix calculation $[Q_5^b, N]$ is equivalent (up to a multiplicative factor)
to the $SU(3)_A$ chiral transformation which we have found in our previous papers:
Eqs.~(11) and (13) in Ref. \cite{Chen:2009sf},
lead to
\begin{eqnarray}
\label{def:N18} [Q_5^a, N_{(6,3)}] &=&
\gamma_5 \left(\left({\rm {\bf D}_{(8)}^{a} + {2\over3} {\bf F}_{(8)}^{a}} \right) N_{(6,3)}
+ {2\over\sqrt3}{\rm {\bf T}_{(8/10)}^a \Delta_{(6,3)}} \right) \, , \\
\label{def:Delta18}
[Q_5^a , \Delta_{(6,3)} ] &=&
\gamma_5 \left({2\over\sqrt3} {\rm {\bf T}^{\dagger a}}_{(8/10)} N_{(6,3)} +
{1\over3} {\bf F}_{(10)}^{a} \Delta_{(6,3)} \right)
 \, , \\
\nonumber [Q_5^a, N_{(3,\bar3)}] &=& ~~\gamma_5 {\rm {\bf D}}^{a} N_{(3,\bar3)}  \, ,
\\ \nonumber [Q_5^a, N_{(\bar3,3)}] &=& - \gamma_5 {\rm {\bf D}}^{a} N_{(\bar3,3)} \, .
\end{eqnarray}
These $SU(3)$-spurion matrices ${\bf T}^{a}$ (sometimes we use
${\bf T}^{a}_{10/8}$) and ${\bf F}_{(10)}^{a}$ have the following
properties
\begin{eqnarray}
{\rm {\bf F}}_{(10)}^{a} &=& - \, i \, f^{abc}{\bf T}_{10/8}^{b
\dagger} {\bf T}_{10/8}^{c}  \, ,
\nonumber \\
{\bf T}_{10/8}^{a} {\bf T}_{10/8}^{a \dagger} &=&  \, \frac52 \times
{\mathbf 1}_{8\times8} \, ,
\nonumber \\
{\bf T}_{10/8}^{a \dagger} {\bf T}_{10/8}^{a} &=&  \, 2 \times {\mathbf
1}_{10\times10} \, . \label{e:app32c}
\end{eqnarray}
The octet generators $\left({\rm {\bf D}_{(8)}^{a} + {2\over3} {\bf F}_{(8)}^{a}} \right)$,
the transition matrices ${\bf T}_{10/8}^{a}$ and the decuplet
generators ${\rm {\bf F}}_{(10)}^{a}$ are listed in
Appendices~A1., A.2 and A.3, respectively, of Ref. \cite{Chen:2009sf}.

\subsubsection{Chiral $SU_L(2) \times SU_R(2)$ Commutators}

For the chiral $SU_L(2) \times SU_R(2)$ subgroup, i.e. for $a=1,2,3$, this
leads to
\begin{eqnarray}
\nonumber [Q_5^a, N_{(1,\frac12)}] &=& \gamma_5 \left({5\over3}
\frac{\tau^a}{2} N_{(1,\frac12)} +
\frac{2}{\sqrt{3}} T^a \Delta_{(1,\frac12)} \right) \, , \\
\label{def:Delta18}
[ Q_5^a , \Delta_{(1,\frac12)} ] &=&
\gamma_5 \left({2\over\sqrt3} {{T}^{\dagger a}} N_{(1,\frac12)} +
{1\over3} {t}_{(3/2)}^{a} \Delta_{(1,\frac12)} \right)  \, , \\
\nonumber [Q_5^a, N_{(\frac12,0)}] &=& \gamma_5 \frac{\tau^a}{2} N_{(\frac12,0)} \, ,
\\ \nonumber [Q_5^a, N_{(0,\frac12)}] &=& - \gamma_5 \frac{\tau^a}{2} N_{(0,\frac12)} \, ,
\end{eqnarray}
where $a=1,2,3$,
${\bf t}_{(\frac32)}^i$ are the isospin-$\frac32$ generators
of the SU(2) group and ${\bf T}^i$ are the so-called iso-spurion
($4\times2$) matrices, see Appendix B of Ref. \cite{Nagata:2008xf}.

Consequently,
\begin{eqnarray}
\nonumber [Q_5^a, \bar N_{(1,\frac12)} N_{(1,\frac12)}] &=&
\frac{5}{3} {\bar N}_{(1,\frac12)} \gamma_5 \tau^a N_{(1,\frac12)} \, \\
&+&
\frac{2}{\sqrt{3}} \left({\bar N}_{(1,\frac12)} \gamma_5 T^a \Delta_{(1,\frac12)}
+ {\bar \Delta}_{(1,\frac12)} \gamma_5 T^{\dagger a} N_{(1,\frac12)} \right)\, ,\\
\nonumber [Q_5^a, \bar N_{(\frac12,0)} N_{(\frac12,0)}] &=&
\bar N_{(\frac12,0)} \gamma_5 \tau^a N_{(\frac12,0)} \, ,
\\ \nonumber [Q_5^a, \bar N_{(0,\frac12)} N_{(0,\frac12)}] &=& -
\bar N_{(0,\frac12)} \gamma_5 \tau^a N_{(0,\frac12)} \, ,
\end{eqnarray}
and similarly for the $\Delta$-field commutator
\begin{eqnarray}
\nonumber [Q_5^a, \bar \Delta_{(1,\frac12)} \Delta_{(1,\frac12)}] &=&
\frac{2}{3} {\bar \Delta}_{(1,\frac12)} \gamma_5 {t}_{(3/2)}^{a} \Delta_{(1,\frac12)} \, \\
&+&
\frac{2}{\sqrt{3}} \left({\bar N}_{(1,\frac12)} \gamma_5 T^a \Delta_{(1,\frac12)}
+ {\bar \Delta}_{(1,\frac12)} \gamma_5 T^{\dagger a} N_{(1,\frac12)} \right)\, .
\end{eqnarray}

\subsubsection{Chiral $SU_L(2) \times SU_R(2)$ Double Commutators}

Here we have
\begin{eqnarray}
\nonumber [Q_5^b, [Q_5^a, \bar N_{(1,\frac12)} N_{(1,\frac12)}]] &=&
\left({5\over3}\right)^2 \delta^{ab}  \bar N_{(1,\frac12)} N_{(1,\frac12)}  \\
&+&
\nonumber
\left({5\over3}\right) \frac{1}{\sqrt{3}} \left({\bar N}_{(1,\frac12)} \tau^b T^a
\Delta_{(1,\frac12)}
+ {\bar \Delta}_{(1,\frac12)} T^{\dagger a} \tau^b N_{(1,\frac12)} \right)  \\
&+&
\nonumber
[Q_5^b, \frac{2}{\sqrt{3}} \left({\bar N}_{(1,\frac12)} \gamma_5 T^a \Delta_{(1,\frac12)}
+ {\bar \Delta}_{(1,\frac12)} \gamma_5 T^{\dagger a} N_{(1,\frac12)} \right)] \\
&=&
\nonumber
\left(\frac{25+16}{9}\right) \delta^{ab}  \bar N_{(1,\frac12)} N_{(1,\frac12)} \\
\label{e:doublecommN}
&+& \left({4\over3}\right)
{\bar \Delta}_{(1,\frac12)} \left(\frac{3}{2} \delta^{ab} - \frac{1}{3} \left\{{t}_{(3/2)}^{a},
{t}_{(3/2)}^{b} \right\} \right) \Delta_{(1,\frac12)} + \ldots\, ,
\end{eqnarray}
where $\ldots$ stand for the off-diagonal terms, such as
${\bar N}_{(1,\frac12)} (\ldots) \Delta_{(1,\frac12)}$, and their Hermitian conjugates.
Similarly for the $\Delta$-field double commutator
\begin{eqnarray}
\nonumber [Q_5^b, [Q_5^a, \bar \Delta_{(1,\frac12)} \Delta_{(1,\frac12)}]]
&=&
\nonumber
\left(\frac{16}{9}\right) \delta^{ab}  \bar N_{(1,\frac12)} N_{(1,\frac12)}
+
2 \delta^{ab} {\bar \Delta}_{(1,\frac12)} \Delta_{(1,\frac12)} \\
\label{e:doublecommDelta}
&-& \left(\frac{2}{9}\right) {\bar \Delta}_{(1,\frac12)} \left\{{t}_{(3/2)}^{a},
{t}_{(3/2)}^{b} \right\} \Delta_{(1,\frac12)}  + \ldots\, ,
\end{eqnarray}
where $\ldots$ again stand for the off-diagonal terms.
The $(\frac12,0)$ and $(0,\frac12)$ chiral multiplets double commutators are much simpler
\begin{eqnarray}
\label{e:doublecommN3} [Q_5^b, [Q_5^a, \bar N_{(\frac12,0)} N_{(\frac12,0)}]]
&=& \delta^{ab} \bar N_{(\frac12,0)} N_{(\frac12,0)} \, ,\\
\label{e:doublecommN03} [Q_5^b, [Q_5^a, \bar N_{(0,\frac12)} N_{(0,\frac12)}]]
&=& \delta^{ab} \bar N_{(0,\frac12)} N_{(0,\frac12)} \, .
\end{eqnarray}
Finally, we contract these Eqs. (\ref{e:doublecommN}),
(\ref{e:doublecommDelta}),(\ref{e:doublecommN3}),(\ref{e:doublecommN03})
with $\frac13 \delta^{ab}$ (where summation over repeated indices is understood)
to find:
\begin{eqnarray}
\frac13 \delta^{ab} [Q_5^b, [Q_5^a, \bar N_{(1,\frac12)} N_{(1,\frac12)}]] &=&
\label{e:NSigma112}
\left(\frac{41}{9}\right) \bar N_{(1,\frac12)} N_{(1,\frac12)}
+ \left({8\over9}\right) {\bar \Delta}_{(1,\frac12)} \Delta_{(1,\frac12)} + \ldots \, ,
\end{eqnarray}
and similarly
\begin{eqnarray}
\frac13 \delta^{ab} [Q_5^b, [Q_5^a, \bar \Delta_{(1,\frac12)} \Delta_{(1,\frac12)}]]
&=&
\label{e:DSigma112}
\left(\frac{16}{9}\right) \bar N_{(1,\frac12)} N_{(1,\frac12)}
+ \left({13\over9}\right)
{\bar \Delta}_{(1,\frac12)} \Delta_{(1,\frac12)}  + \ldots  \, ,
\end{eqnarray}
where we have used the identity
${t}_{(3/2)}^{a} {t}_{(3/2)}^{a} = {15\over4} {\bf 1}_{4\times4}$.
\begin{eqnarray}
\label{e:NSigma120}
\frac13 \delta^{ab} [Q_5^b, [Q_5^a, \bar N_{(\frac12,0)} N_{(\frac12,0)}]]
&=& \bar N_{(\frac12,0)} N_{(\frac12,0)} \, , \\
\label{e:NSigma012}
\frac13 \delta^{ab} [Q_5^b, [Q_5^a, \bar N_{(0,\frac12)} N_{(0,\frac12)}]]
&=& \bar N_{(0,\frac12)} N_{(0,\frac12)} \, .
\end{eqnarray}

\subsubsection{Chirally Mixed Double Commutators}

As shown in our previous papers, the physical nucleon field contains several chiral
multiplet components:
\begin{eqnarray}
| N \rangle &=& \sin \theta | (6,3) \rangle +
\cos \theta (\cos \varphi | (3,\bar3) \rangle + \sin \varphi | (\bar3,3) \rangle)\, .
\end{eqnarray}
Use the identities
\begin{eqnarray}
N_{(1,\frac12)} | N(p) \rangle &=& N_{(1,\frac12)} \left(\sin \theta | (6,3) \rangle +
\cos \theta (\cos \varphi | (3,\bar3) \rangle + \sin \varphi | (\bar3,3) \rangle) \right)
\nonumber    \\
&=&
u(p)_{(1,\frac12)} \sin \theta  \, ,\\
N_{(\frac12,0)} | N(p) \rangle &=& N_{(\frac12,0)} \left(\sin \theta | (6,3) \rangle +
\cos \theta (\cos \varphi | (3,\bar3) \rangle + \sin \varphi | (\bar3,3) \rangle) \right)
\nonumber \\
&=&
u(p)_{(\frac12,0)} \cos \theta \cos \varphi  \, ,\\
N_{(0,\frac12)} | N(p) \rangle &=& N_{(0,\frac12)} \left(\sin \theta | (6,3) \rangle +
\cos \theta (\cos \varphi | (3,\bar3) \rangle + \sin \varphi | (\bar3,3) \rangle) \right)
\nonumber \\
&=&
u(p)_{(0,\frac12)} \cos \theta \sin \varphi \, ,
\end{eqnarray}
and the Dirac conjugate
\begin{eqnarray}
\langle N(p) | {\bar N}_{(1,\frac12)}  &=& \left(\sin \theta \langle  (6,3)| +
\cos \theta (\cos \varphi \langle (3,\bar3) | + \sin \varphi \langle (\bar3,3) |) \right)
{\bar N}_{(1,\frac12)}
\nonumber    \\
&=&
{\bar u}(p)_{(1,\frac12)} \sin \theta \, ,\\
\langle N(p) | {\bar N}_{(\frac12,0)}  &=& \left(\sin \theta \langle  (6,3)| +
\cos \theta (\cos \varphi \langle (3,\bar3) | + \sin \varphi \langle (\bar3,3) |) \right)
{\bar N}_{(\frac12,0)}
\nonumber \\
&=&
{\bar u}(p)_{(\frac12,0)} \cos \theta \cos \varphi \, ,\\
\langle N(p) | {\bar N}_{(0,\frac12)}  &=& \left(\sin \theta \langle  (6,3)| +
\cos \theta (\cos \varphi \langle (3,\bar3) | + \sin \varphi \langle (\bar3,3) |) \right)
{\bar N}_{(0,\frac12)}
\nonumber \\
&=&
{\bar u}(p)_{(0,\frac12)} \cos \theta \sin \varphi \, ,
\end{eqnarray}
which leads to
\begin{eqnarray}
\langle N(p) | {\bar N}_{(1,\frac12)} N_{(1,\frac12)} | N(p) \rangle
&=&
{\bar u}(p)_{(1,\frac12)} u(p)_{(1,\frac12)} \sin^2 \theta = \frac{E_p}{m} \sin^2 \theta \, ,
\\
\langle N(p) | {\bar N}_{(\frac12,0)}  N_{(\frac12,0)} | N(p) \rangle
&=&
{\bar u}(p)_{(\frac12,0)} u(p)_{(\frac12,0)} \cos^2 \theta \cos^2 \varphi
= \frac{E_p}{m} \cos^2 \theta \cos^2 \varphi  \, ,\\
\langle N(p) | {\bar N}_{(0,\frac12)} N_{(0,\frac12)} | N(p) \rangle
&=&
{\bar u}(p)_{(0,\frac12)} u(p)_{(0,\frac12)} \cos^2 \theta \sin^2 \varphi
= \frac{E_p}{m} \cos^2 \theta \sin^2 \varphi \, ,
\end{eqnarray}
which in the $p \to 0$ limit implies
\begin{eqnarray}
\lim_{p \to 0} \langle N(p) | {\bar N}_{(1,\frac12)} N_{(1,\frac12)} | N(p) \rangle
&=& \sin^2 \theta  \, ,\\
\lim_{p \to 0} \langle N(p) | {\bar N}_{(\frac12,0)}  N_{(\frac12,0)} | N(p) \rangle
&=& \cos^2 \theta \cos^2 \varphi  \, ,\\
\lim_{p \to 0} \langle N(p) | {\bar N}_{(0,\frac12)} N_{(0,\frac12)} | N(p) \rangle
&=& \cos^2 \theta \sin^2 \varphi \, .
\end{eqnarray}
Take the definition
\begin{eqnarray}
\Sigma_{\pi N}^{\alpha} &=&
\frac13 \delta^{ab}
\langle N_{\alpha} | [ Q_5^a, [ Q_5^b, \mathcal{H}_{\rm \chi SB}(0) ] ]
| N_{\alpha} \rangle
\, ,
\end{eqnarray}
where $\alpha$ is the chiral representation, and evaluate it using Eqs. (\ref{e:NSigma112}),
(\ref{e:DSigma112}), (\ref{e:NSigma120}), (\ref{e:NSigma012})
\begin{eqnarray}
\nonumber
\Sigma_{\pi N}^{(1,\frac12)}
&=& \left( \frac{41}{9}  m_N^{(1,\frac12)} + \frac{16}{9} m_{\Delta}^{(1,\frac12)} \right) \, ,
\\
\Sigma_{\pi N}^{(\frac12,0)}
&=& m_N^{(\frac12,0)}  \, ,\\
\nonumber
\Sigma_{\pi N}^{(0,\frac12)}
&=& m_N^{(0,\frac12)}  \, ,
\end{eqnarray}
thus find
\begin{eqnarray}
\nonumber
\Sigma_{\pi N} &=&
\frac13 \delta^{ab} \langle N | [ Q_5^a, [ Q_5^b, \mathcal{H}_{\rm \chi SB}(0) ] ]
| N \rangle \\
\nonumber
&=&  \sin^2 \theta ~\Sigma_{\pi N}^{(1,\frac12)} +
\cos^2 \theta \left(\cos^2 \varphi ~\Sigma_{\pi N}^{(\frac12,0)} + \sin^2 \varphi
~\Sigma_{\pi N}^{(0,\frac12)} \right) \\
\label{e:Sigmafinal} &=&  \sin^2 \theta ~\left( \frac{41}{9}  m_N^{(1,\frac12)}
+ \frac{16}{9} m_{\Delta}^{(1,\frac12)} \right)  +
\cos^2 \theta \left(\cos^2 \varphi ~m_N^{(\frac12,0)} + \sin^2 \varphi
~m_{N}^{(0,\frac12)} \right) \, ,
\end{eqnarray}
which is our basic result here.

Assuming that all three chiral components have the same ``current'' nucleon mass
$M_{N}^0 = M_{(6,3)}^0 = m_N^{(1,\frac12)} = m_{\Delta}^{(1,\frac12)}
= M_{(3,\bar3)}^0 = m_N^{(\frac12,0)} = M_{(\bar3,3)}^0 = m_{N}^{(0,\frac12)}$
one finds finally
\begin{eqnarray}
\Sigma_{\pi N} &=&
\left( \frac{57}{9} \sin^2 \theta ~+ \cos^2 \theta  \right) M_{N}^0
\nonumber \\
&=&
\left(1+ \frac{16}{3} \sin^2 \theta \right) M_{N}^0 \, .
\label{e:Sigma_final}
\end{eqnarray}
Note that the factor $\left(1+ \frac{16}{3} \sin^2 \theta \right)$ in front of
the current nucleon mass is always larger than unity (for real values of the mixing
angle $\theta$).

\subsection{Comparison with experiment}
\label{ss:experiment}

As most ``measurements'' of $\Sigma_{\pi N}$ have yielded values ranging from
55 MeV to 75 MeV~\footnote{Phenomenologically, the sigma term is related to
the $\pi N$ scattering amplitude at a certain non-physical kinematical point,
(at the so-called Cheng-Dashen point $t = + 2 m_{\pi}^2$).
Its extracted value is typically in the range $\Sigma_{CD}$ = 70-90 MeV.
After the corrections for the finite value of $t$ are taken into account,
which roughly amount to -15 MeV, one obtains $\Sigma_{\pi N}$  = 55-75 MeV.},
that are substantially larger than the naively expected 25 MeV, it has consequently appeared that
the $s \bar s$ content of the nucleon must be (very) large.

The nucleon current mass is
$M_{N}^{0} = 3 {\bar m}_{q}^{0}$ = ${3 \over 2} \left(m_{u}^{0} + m_{d}^{0}\right) \simeq$
14.4 MeV, i.e. ${1 \over 2} \left(m_{u}^{0} + m_{d}^{0}\right) \simeq 4.79$ MeV in PDG2012~\cite{Beringer:1900zz}.
We note that here $m_u^{0} = 2.3 \times 1.35$ MeV and $m_d^{0} = 4.8 \times 1.35$ MeV,
where 1.35 is the rescaling factor due to the change of the energy scale from 2 GeV
down to 1 GeV~\cite{Beringer:1900zz}, yielding
${1 \over 2} \left(m_{u}^{0} + m_{d}^{0}\right) \simeq 4.79$ MeV, substantially lower
than 7.6 MeV in PDG1998~\cite{Caso:1998tx}.

The
constraint on the mixing angle $\theta$ by the experimental values of the axial couplings
has been discussed in Sec.~\ref{sect:Phenomenology}, which gives $\theta = 50.7^{\rm o}$ and $\varphi = 66.1^{\rm o}$.
Inserting these values into Eq. (\ref{e:Sigma_final}), one finds
$\Sigma_{\pi N}=$60.3 MeV.

The $\Sigma$ operator, Eq.(\ref{e:Sigma_def}), is often identified with the
chiral symmetry breaking ($\chi$SB) Hamiltonian itself. In
Eq. 
(\ref{e:Nucleon_mass})
the nucleon $\Sigma$ term is  a measure of the $\chi$SB in the nucleon.
In such a case it equals the shift of the nucleon mass $\delta M$ due to the
$\chi$SB terms in the Hamiltonian.
This reasoning underlies the standard interpretation
of the nucleon $\Sigma$ term as being a measure of
the strangeness content of the nucleon Ref. \cite{kj87}.

A large value of $\Sigma_{\pi N}$, such as 65 MeV, has often been interpreted as
a sign of a substantial $s {\bar s}$ content of the nucleon.
We have shown that in the chiral-mixing approx 
large values of $\Sigma_{\pi N}$ can be obtained  without any strangeness
degrees of freedom in the nucleon as a natural consequence of the
rather substantial chiral $[({\bf 6},{\bf
3})\oplus({\bf 3},{\bf 6})]$ multiplet component in the nucleon
field.

\section{The kaon-nucleon $\Sigma_{K N}$ term}
\label{sect:K_Sigma_term}


\subsection{SU(3) Quark level}

To calculate the kaon-nucleon $\Sigma_{K N}$ term, we use the $\Sigma^{ab}$ operator defined
as the double commutator
\begin{eqnarray}
\label{e:Sigma_KN_def}
\Sigma^{ab} &=&
[ Q_5^a, [ Q_5^b, {H}_{\rm \chi SB}] ]  \, ,
\end{eqnarray}
of the axial charges $Q^{a}_{5}$ and the chiral symmetry breaking Hamiltonian
${H}_{\chi SB}$~\footnote{For normalization
and notational conventions see Ref.~\cite{Dmitrasinovic:1999mf}.}.
It was introduced by Dashen~\cite{dash69}
as a way of separating out the explicit chiral $SU_L(3) \times SU_R(3)$ symmetry
breaking part ${H}_{\chi SB}$ from the total Hamiltonian.

Its (diagonal) nucleon matrix element $\Sigma_{K N} = \frac14 \sum_{a=4}^{7}
\langle N | \Sigma^{aa} | N \rangle$
is due to the (explicit) chiral symmetry
breaking current quark masses~\cite{dash69,reya74}.
Then the kaon-nucleon $\Sigma$-terms are
\begin{eqnarray}
\nonumber 
\Sigma^{44}  &=& \left(
\begin{array}{ccc}
 2 \left(m_s+m_u\right) & 0 & 0 \\
 0 & 0 & 0 \\
 0 & 0 & 2 \left(m_s+m_u\right)
\end{array}
\right)  \, ,\\
\nonumber
\Sigma^{55}  &=& \left(
\begin{array}{ccc}
 2 \left(m_s+m_u\right) & 0 & 0 \\
 0 & 0 & 0 \\
 0 & 0 & 2 \left(m_s+m_u\right)
\end{array}
\right)  \, ,\\
\nonumber
\Sigma^{66}  &=& \left(
\begin{array}{ccc}
 0 & 0 & 0 \\
 0 & 2 \left(m_d+m_s\right) & 0 \\
 0 & 0 & 2 \left(m_d+m_s\right)
\end{array}
\right) \nonumber  \, , \\
\Sigma^{77}  &=& \left(
\begin{array}{ccc}
 0 & 0 & 0 \\
 0 & 2 \left(m_d+m_s\right) & 0 \\
 0 & 0 & 2 \left(m_d+m_s\right)
\end{array}
\right) \, .
\nonumber
\end{eqnarray}
Summing them up and dividing by 4, we find
\begin{eqnarray}
\nonumber 
\Sigma_{K N} &=& \frac14 \sum_{a=4}^{7}
\Sigma^{aa} \\
\nonumber
  &=& \left(
\begin{array}{ccc}
 m_s+m_u & 0 & 0 \\
 0 & m_d+m_s & 0 \\
 0 & 0 & m_d+2 m_s+m_u
\end{array}
\right)  \, .
\nonumber
\end{eqnarray}
If we assume that $m_u = m_d$, then
\begin{eqnarray*}
\Sigma_{K N} = (m_s + m_{u/d}) \, .
\end{eqnarray*}


\subsection{SU(3) Hadron level}

Double commutator of the axial charges $Q^{a}_{5}$ for $a=4,5,6,7$ and the current/bare
nucleon
mass Hamiltonian ${H}_{\chi SB}$ (also) gives the kaon $\Sigma$ term operator
\begin{eqnarray*}
\Sigma_{K} = \frac14 \sum_{a=4,5,6,7} [ Q_5^a, [ Q_5^a, {H}_{\rm \chi SB}]] \, .
\end{eqnarray*}
Kaon-nucleon $\Sigma_{K N}$ term - matrix element $\Sigma_{K N} =
\frac14 \sum_{a=4,5,6,7} \langle N | [ Q_5^a, [ Q_5^a, \mathcal{H}_{\rm \chi SB}(0) ] ]
| N \rangle = \langle N | \Sigma_{K} | N \rangle $ - is also a chiral mixture:
\begin{eqnarray*}
\Sigma_{K N} &=& \sin^2 \theta ~
\Sigma_{K N(6,3)} \\
\nonumber &+&
 \cos^2 \theta \left(\cos^2 \varphi ~ \Sigma_{K N(3,\bar3)} + \sin^2 \varphi
~ \Sigma_{K N(\bar3,3)} \right) \, .
\end{eqnarray*}
Thus, we need three double commutators:
\subsubsection{The $({\bf 6},{\bf 3})$ and $({\bf 3},{\bf 6})$ chiral multiplets}
\begin{eqnarray*}
&& {1\over4} \sum_{i=4}^7 ( [ Q_5^i, [ Q_5^i, {\bar N}_{(6,3)} M_{N} N_{(6,3)}] ] \\
&=& {1\over4} \bar N_{(6,3)} \left(\begin{array}{cc}
{70\over9}m_u + {41\over9}m_d + {5\over3}m_s & 0
\\ 0 & {41\over9}m_u + {70\over9}m_d + {5\over3}m_s
\end{array}\right) N_{(6,3)} \, ,
\\ \nonumber
&& {1\over4} \sum_{i=4}^7 ( [ Q_5^i, [ Q_5^i, {\bar \Delta}_{(6,3)} M_{\Delta} \Delta_{(6,3)}] ] \\
&=& {1\over4} {\bar N}_{(6,3)} \left(\begin{array}{cc}
{20\over9}m_u + {4\over9}m_d + {4\over3}m_s & 0
\\ 0 & {4\over9}m_u + {20\over9}m_d + {4\over3}m_s
\end{array}\right) N_{(6,3)}  \, ,
\end{eqnarray*}
thus leading to
\begin{eqnarray*}
\nonumber
\Sigma_{K N(6,3)} &=&  {1\over4} \left(10m_u + 5m_d + 3m_s \right)  \, .
\end{eqnarray*}

\subsubsection{The $(\mathbf{\bar 3},\mathbf{3})$ and $(\mathbf{3},\mathbf{\bar 3})$ chiral multiplets}
\begin{eqnarray*}
\nonumber
&& {1\over4} \sum_{i=4}^7
[ Q_5^i, [ Q_5^i, {\bar N}_{(\bar3,3)} M_{N} N_{(\bar3,3)}] ] \\
&=& {1\over4} \bar N_{(\bar3,3)} \left(\begin{array}{cc}
{26\over3}m_u + {11\over3}m_d + {5\over3}m_s & 0
\\ 0 & {11\over9}m_u + {26\over9}m_d + {5\over3}m_s
\end{array}\right) N_{(\bar3,3)} \, ,\\
&& {1\over4} \sum_{i=4}^7
( [ Q_5^i, [ Q_5^i, {\bar \Lambda}_{(\bar3,3)} M_{\Lambda} \Lambda_{(\bar3,3)}] ] \\
&=& {1\over4} \bar N_{(\bar3,3)} \left(\begin{array}{cc}
{4\over3}m_u + {4\over3}m_d + {4\over3}m_s & 0
\\ 0 & {4\over3}m_u + {4\over3}m_d + {4\over3}m_s
\end{array}\right) N_{(\bar3,3)}   \, ,
\nonumber
\end{eqnarray*}
thus leading to
\begin{eqnarray*}
\Sigma_{K N(\bar3,3)} = {1\over4} \left(10m_u + 5m_d + 3m_s \right) \, .
\end{eqnarray*}

\subsubsection{The $({\bf 8},{\bf 1})$ and $({\bf 1},{\bf 8})$ chiral multiplets}

\begin{eqnarray*}
&& {1\over4} \sum_{i=4}^7
( [ Q_5^i, [ Q_5^i, {\bar N}_{(8,1)} M_{N} N_{(8,1)}] ] \\
&=&
{1\over4} \bar N_{(8,1)} \left(\begin{array}{cc}
10m_u + 5m_d + 3m_s & 0
\\ 0 & 5m_u + 10m_d + 3m_s
\end{array}\right) N_{(8,1)}  \, ,
\end{eqnarray*}
thus leading to
\begin{eqnarray*}
\Sigma_{K N(8,1)} = {1\over4} \left(10m_u + 5m_d + 3m_s \right) \, .
\end{eqnarray*}


\subsection{Numerical results}

One can see that the kaon-nucleon $\Sigma_{K N}$ terms are identical in these three
chiral multiplets, $\Sigma_{K N(6,3)} = \Sigma_{K N(\bar3,3)} = \Sigma_{K N(8,1)}$,
so that the $\Sigma_{K N}$ term of their admixture also equals the same number:
\begin{eqnarray*}
\Sigma_{K N} &=&
\Sigma_{K N(6,3)} \left(\sin^2 \theta +
 \cos^2 \theta \left(\cos^2 \varphi  + \sin^2 \varphi \right) \right)\\
 &=& \Sigma_{K N(6,3)}  = \Sigma_{K N(\bar3,3)} = \Sigma_{K N(8,1)} \\
 &=&  {1\over4} \left(10m_u + 5m_d + 3m_s \right)  \, .
\end{eqnarray*}
The 2012 edition of the Particle Data Group, 
Ref.~\cite{Beringer:1900zz} has $m_u^{0} = 2.3 \times 1.35$ MeV and $m_d^{0} = 4.8 \times 1.35$ MeV,
i.e., ${1 \over 2} \left(m_{u}^{0} + m_{d}^{0}\right) = 4.79$ MeV
and $m_s^{0} = (93.5 \pm 2.5) \times 1.35 = (126.225 \pm 3.375)$ MeV, yielding
\begin{eqnarray*}
\Sigma_{K N} = 111 ~{\rm MeV} \, .
\end{eqnarray*}
Note that these values are substantially lower than before,
see e.g. the PDG1998 values, Ref.~\cite{Caso:1998tx}.

One would like to compare this value with the ``experimental'' one.
In this case, the status of ``experimental'' $\Sigma_{KN}$
is even worse than that of  ``experimental'' $\Sigma_{\pi N}$: the
kaon-nucleon scattering data are nowhere near of pion-nucleon ones
in terms of overall quality, abundance, kinematic range, precision and accuracy.

Only some very old ``experimental'' estimates are available:
1) $\Sigma_{KN} \simeq$ 170 MeV from 1970, Ref.~\cite{Hippel:1970},
2) $\Sigma_{KN}$ = -370 $\pm$ 110 MeV from 1972, Ref.~\cite{Thompson:1971},
$\Sigma_{KN} \simeq$ 170 MeV, from 1973, Ref.~\cite{Nasrallah:1973},
3) $\Sigma_{KN}$ = 540 $\pm$ 160 MeV, from 1973, Ref.~\cite{Reya:1973},
4) $\Sigma_{KN}$ = 246 MeV, from 1976, Ref.~\cite{Cheng:1976}.
The most recent reviews, Refs.~\cite{Olin:2001,kj87} have calculated
$\Sigma_{K N}$ at zero strangeness content $y_N =0$ of the nucleon as
being 170 MeV, but with the 1987 values of current quark masses.
Their formulae translate to the value of 110 MeV with
the 2012 values of masses, Ref.~\cite{Beringer:1900zz}.
This is (very) close to our predicted value (111 MeV) of the same quantity.


\section{Summary and Conclusions}
\label{sec:summary}

In this paper we have calculated the pion-nucleon $\Sigma_{\pi N}$ term in the chiral
mixing approach, first with two light ($u,d$) flavors, and then we extended it to the case
with three light ($u,d,s$) flavors, i.e., to $SU_L(3) \times SU_R(3)$ multiplet mixing,
which we then used to calculate the kaon-nucleon sigma term $\Sigma_{K N}$.
We based our calculations on the chiral mixing formalism and the phenomenology developed
previously
in Refs. \cite{Chen:2008qv,Chen:2009sf,Chen:2010ba,Dmitrasinovic:2014aya,Dmitrasinovic:2014eia}.

The physical significance of our present work is that it shows that there is no need to
introduce $s {\bar s}$ components in addition to the three-quark ``core'', so as to agree
with the observed values of the pion-nucleon $\Sigma$ term, the baryon axial couplings, and
the nucleon magnetic moments: the phenomenologically necessary
$[(\mathbf{6},\mathbf{3}) \oplus (\mathbf{3},\mathbf{6})]$ chiral component and the
$[(\bar{\mathbf{3}},\mathbf{3})\oplus(\mathbf{3},\bar{\mathbf{3}})]$
``mirror'' component exist as bi-local three-quark fields, Ref. \cite{Dmitrasinovic:2011yf,Chen:2012vs}.
Thus, we have shown that there is no need for ``meson cloud'', or (non-exotic) ``pentaquark''
components in the Fock expansion of the baryon wave function, to explain (at least)
the axial currents, magnetic moments and the pion-nucleon $\Sigma$ term, contrary to
established opinion, Ref. \cite{Donoghue:1985bu}. This goes to show that the algebraic
complexity of three Dirac quark fields is such that it can mimick the presence of
$q {\bar q}$ pairs, at least in certain observables. For us this was a surprise.

The present formalism and phenomenology can be used to attack other outstanding
issues of baryon chiral dynamics: the hyperon radiative decays, for example, have
been a long-standing unsolved problem.


\section*{Acknowledgments}

The work of H.X.C is supported by the National Natural Science Foundation of China under
Grant No. 11475015. The work of V.D. was supported by the Serbian
Ministry of Science and Technological Development under grant numbers OI 171037 and III 41011.
AH is supported in part by the Grant-in-Aid  for Science Research (C) 26400273.


\appendix

\section{Isospin-$\frac32$ generators}
\label{App:Isospin matrices}

From Ref. \cite{Nagata:2008xf} we have
\begin{eqnarray}
\delta_5^{a_3} \phi_{\frac32,\frac32}^{\mu} &=& i\gamma_5 a_3
\phi_{\frac32,\frac32}^{\mu},
\label{e:chiral32d}\\
\delta_5^{a_3}\left(\begin{array}{c} \phi_{\frac12,\frac12}^{\mu} \\
\phi_{\frac32,\frac12}^{\mu}\end{array}\right)
&=& i \gamma_5 a_3 \left(\begin{array}{cc} \frac53 & \frac{4\sqrt{2}}{3}\\
\frac{4\sqrt{2}}{3} & \frac13 \end{array}\right)
\left(\begin{array}{c} \phi_{\frac12,\frac12}^{\mu}\\
\phi_{\frac32,\frac12}^{\mu} \end{array}\right),
\label{e:chiral32f}
\end{eqnarray}
with the familiar (``$SU_{\rm FS}(6)$") value $\frac53$ for its
``nucleon" component $N^{\mu}$. In order to read off the value of
$g_A$, it is convenient to express this 
as
\begin{eqnarray}
\delta_5^{\vec{a}}\Delta^{\mu} &=& \, i \,\gamma_5\, \left(
\frac13 \, \inner{t_{(\frac32)}}{a} \, \Delta^{\mu} +
\frac{2}{\sqrt{3}} \, \inner{a}{T^{\dagger}} \,N^{\mu}\right),
\label{e:chiral32c}
\end{eqnarray}
where ${\bf t}_{(\frac32)}^i$ are the isospin-$\frac32$ generators
of the SU(2) group and ${\bf T}^i$ are the so-called iso-spurion
($4\times2$) matrices, see Appendix B of Ref. \cite{Nagata:2008xf}.

The $t_{(\frac32)}^i$ are defined as
\begin{eqnarray}
{\bf t}_{(\frac32)}^1 &=& \left(\begin{array}{cccc}
0 & \frac{{\sqrt{3}}}{2} & 0 & 0 \\
\frac{{\sqrt{3}}}{2} & 0 & 1 & 0 \\
0 & 1 & 0 & \frac{{\sqrt{3}}}{2} \\
0 & 0 & \frac{{\sqrt{3}}}{2} & 0 \end{array}
\right)  \nonumber \, , \\
{\bf t}_{(\frac32)}^2 &=& i \left(\begin{array}{cccc}
0 & -\frac{{\sqrt{3}}}{2} & 0 & 0 \\
\frac{{\sqrt{3}}}{2} & 0 & - 1 & 0 \\
0 & 1 & 0 & - \frac{{\sqrt{3}}}{2} \\
0 & 0 & \frac{{\sqrt{3}}}{2} & 0 \end{array}
\right)  \nonumber \, , \\
{\bf t}_{(\frac32)}^3 &=&  \left(\begin{array}{cccc}
\frac{3}{2} & 0 & 0 & 0 \\
0 & \frac{1}{2} & 0 & 0 \\
0 & 0 & -\frac{1}{2} & 0 \\
0 & 0 & 0 & -\frac{3}{2} \end{array} \right) \, , \label{e:appb}
\end{eqnarray}
which leads to the conventional normalization of the SU(2) Casimir
operator.
The ${\bf T}^i$ are defined by
\begin{eqnarray}
{\bf T}^1 &=& \left(\begin{array}{cccc}
- \frac{1}{{\sqrt{2}}} & 0 & \frac{1}{{\sqrt{6}}} & 0 \\
0 & - \frac{1}{{\sqrt{6}}} & 0 & \frac{1}{{\sqrt{2}}}
\end{array}
\right)  \nonumber \, , \\
{\bf T}^2 &=& i \left(\begin{array}{cccc}
- \frac{1}{{\sqrt{2}}} & 0 & - \frac{1}{{\sqrt{6}}} & 0 \\
0 & - \frac{1}{{\sqrt{6}}} & 0 & - \frac{1}{{\sqrt{2}}}\end{array}
\right)  \nonumber \, , \\
{\bf T}^3 &=& \left(\begin{array}{cccc}
0 & {\sqrt{\frac{2}{3}}} & 0 & 0 \\
0 & 0 & {\sqrt{\frac{2}{3}}} & 0 \end{array} \right) \, ,
\label{e:appbc}
\end{eqnarray}
with the properties
\begin{eqnarray}
i \, \inner{t_{(\frac32)}}{a} &=& \frac32 {\bf T}^{i\dagger}\left(
i \inner{\tau}{a}\delta^{ik}\right) {\bf T}^{k} \, =
- \frac32 {\bf T}^{i\dagger}\left(\epsilon^{ijk} {a}^j \right) {\bf T}^{k}  \, , \nonumber \\
{\bf T}^{i} {\bf T}^{k\dagger} &=&  \, P_{\frac32}^{ik} \, .
\label{e:app32b}
\end{eqnarray}

\section{Closure of the chiral $SU_L(3) \times SU_R(3)$ algebra}
\label{sect:SU(3)chiral algebra}

The $SU(3)$ vector charges $Q^a = \int d {\bf x}J_{0}^a (t,{\bf x})$
defined as
\begin{eqnarray}
- 2 \inner{b}{J_{\mu}} &=& \sum_{i} \frac{\partial {\cal
L}}{\partial \partial^{\mu} B_{i}} \delta^{\vec{b}} B_{i} \, ,
\label{e:Noether vec} \
\end{eqnarray}
together with the axial charges $Q_{5}^a = \int d {\bf x}J_{05}^a (t,{\bf x})$ defined as
\begin{eqnarray}
- 2 \inner{a}{J_{\mu 5}} &=& \sum_{i} \frac{\partial {\cal
L}}{\partial \partial^{\mu} B_{i}} \delta_{5}^{\vec{a}} B_{i} \, ,
\label{e:Noether1} \
\end{eqnarray}
ought to close the chiral algebra
\begin{eqnarray}
\left[Q^a , Q^b \right] &=& i f^{abc} Q^c \, ,
\label{e:cc3VVV} \\
\left[Q_{5}^a , Q^b \right] &=& i f^{abc} Q_{5}^c \, ,
\label{e:cc3AVA} \\
\left[Q_{5}^a , Q_{5}^b \right] &=& i f^{abc} Q^c \, ,
\label{e:cc3AAA}
\end{eqnarray}
where $f^{abc}$ are the SU(3) structure constants.
Eqs.~(\ref{e:cc3VVV}) and (\ref{e:cc3AVA}) usually hold
automatically, as a consequence of the canonical (anti)commutation
relations between Dirac baryon fields $B_{i}$, whereas
Eq.~(\ref{e:cc3AAA}) is not trivial for the chiral multiplets that
are different from the $[(8,1) \oplus (1,8)]$, because of the
(nominally) fractional axial charges and the presence of the
off-diagonal components. When taking a matrix element of
Eq.~(\ref{e:cc3AAA}) by baryon states in a certain chiral
representation, the axial charge mixes different flavor states
within the same chiral representation. This is an algebraic version
of the Adler-Weisburger sum rule~\cite{Weinberg:1969hw}. In the
following we shall check and confirm the validity of
Eq.~(\ref{e:cc3AAA}) in the three multiplets of SU(3)$_L
\times$SU(3)$_R$.

\subsection{Closure of the Chiral $SU_L(3) \times SU_R(3)$ Algebra
in the $(8,1) \oplus (1, 8)$ Multiplet} \label{ssect:closure81a}

Due to the absence of fractional coefficients in the $(8,1) \oplus
(1, 8)$ multiplet's axial charge $Q_{5}^a = \int d {\bf x}J_{05}^a (t,{\bf x})$ defined by the current given in
\begin{eqnarray}
{\bf J}_{\mu 5}^{a} &=& \,\overline{N} \gamma_{\mu} \gamma_{5}
{\bf F}_{(8)}^{a} N  \label{e:axi1SU(3)81} \, ,
\end{eqnarray}
the vector charge $Q^a = \int d {\bf x}J_{0}^a (t,{\bf x})$ defined by the current given in
\begin{eqnarray}
{\bf J}_{\mu}^{a} &=& \overline{N} \gamma_{\mu} \, {\bf
F}_{(8)}^{a} \,N   \, , \label{e:vec1SU(3)81} \
\end{eqnarray}
and the axial charge close the chiral
algebra defined by Eqs.~(\ref{e:cc3VVV}), (\ref{e:cc3AVA}) and
(\ref{e:cc3AAA}). The same comments holds for the $(10,1) \oplus (1,
10)$ chiral multiplet for the same reasons as in the example shown
above.

\subsection{Closure of the Chiral $SU_L(3) \times SU_R(3)$ Algebra
in the $(3,\overline{3}) \oplus (\overline{3}, 3)$ Multiplet}
\label{ssect:closure33a}

The vector charge $Q^a = \int d {\bf x}J_{0}^a (t,{\bf x})$ defined
by the current given in
\begin{eqnarray}
{\bf J}_{\mu}^{a} &=& \overline{N} \gamma_{\mu} \, {\bf
F}_{(8)}^{a} \,N  \, , \label{e:vec1SU(3)33} \
\end{eqnarray}
together with the
axial charge $Q_{5}^a = \int d {\bf x}J_{05}^a (t,{\bf x})$ defined
by the current given in
\begin{eqnarray}
{\bf J}_{\mu 5}^{a} &=& \,\overline{N} \gamma_{\mu} \gamma_{5}
\left({\rm {\bf D}}^{a} N + \sqrt{2 \over 3} {\rm \bf
T}^{a\dagger}_{1/8} \Lambda_1 \right)
+ \, \overline{\Lambda}_1 \gamma_{\mu} \gamma_{5} \sqrt{2 \over
3} {\rm \bf  T}^a_{1/8} N  \, , \label{e:axi1SU(3)33} \
\end{eqnarray}
ought to close the
chiral algebra defined by Eqs.~(\ref{e:cc3VVV}), (\ref{e:cc3AVA})
and (\ref{e:cc3AAA}). Eqs.~(\ref{e:cc3VVV}) and (\ref{e:cc3AVA})
hold here, whereas Eq.~(\ref{e:cc3AAA}) is the non-trivial one: the
diagonal $D$ charge of $N$ ($Q_{5D}^a(N)$) axial charge,
\begin{eqnarray}
Q_{5D}^a(N) &=&  ~~~ \int d {\bf x} \, \left(\overline{N} \gamma_{0}
\gamma_{5} \,{\bf D}^{a} \,N \right)
\label{e:offdiag1dSU(3)} \,, \\
Q_{D}^a(N) &=&  ~~~ \int d {\bf x} \, \left(\overline{N} \gamma_{0}
\,{\bf D}^{a} \,N \right) \label{e:offdiag1dNSU(3)} \, ,  \
\end{eqnarray}
lead to
\begin{eqnarray}
\left[Q_{5D}^a(N) , Q_{5D}^b(N) \right] &=& \int d {\bf x}
\left(\overline{N} \gamma_{0} \,\left({\bf D}^{a} {\bf D}^{b} -
{\bf D}^{b} {\bf D}^{a} \right) N \right) \, .
\label{e:ccrAAVS36b} \
\end{eqnarray}
It turns out that the off-diagonal terms in the axial charge
\begin{eqnarray}
Q_{5}^a(N,\Lambda) &=& \int d {\bf x} \, \Bigg( \sqrt{\frac{2}{3}}
\left(\overline{N} \gamma_{0} \gamma_{5} \,{\bf T}^{a \dagger}_{1/8}
\,\Lambda + \overline{\Lambda} \gamma_{0} \gamma_{5} \,{\bf
T}^a_{1/8} \,N \right) \Bigg) \, , \label{e:axioffdiag1Sigma}\
\end{eqnarray}
play a crucial role in the closure of the chiral commutator
Eq.~(\ref{e:cc3AAA}). The additional terms in the commutator add up
to
\begin{eqnarray}
\left[Q_{5}^a(N,\Delta) , Q_{5}^b(N,\Delta) \right] &=&
{\frac{2}{3}} \int d {\bf x}  \overline{N} \gamma_{0} \,\left({\bf
T}^{a\dagger}_{1/8} {\bf T}^{b}_{1/8} - {\bf T}^{b \dagger}_{1/8}
{\bf T}^{a}_{1/8} \right) \,N \, , \label{e:ccrAAVS36a} \
\end{eqnarray}
which provide the ``missing'' factors due to the following
properties of the off-diagonal isospin operators ${\bf T}^{i}_{1/8}$
and ${\bf D}^{i}$ matrices
\begin{eqnarray}
i \,f^{ijk} ({\bf F}_{(8)}^{k}) &=&  ({\bf D}^{i} {\bf D}^{j} - {\bf
D}^{j} {\bf D}^{i}) + {2\over3} ({\bf T}^{i\dagger}_{1/8} {\bf
T}^{j}_{1/8} - {\bf T}^{j\dagger}_{1/8} {\bf T}^{i}_{1/8}) \, .
\label{e:app1bSigma}
\end{eqnarray}
Therefore, the chiral algebra Eqs.~(\ref{e:cc3VVV}),
(\ref{e:cc3AVA}) and (\ref{e:cc3AAA}) close.

\subsection{Closure of the Chiral $SU_L(3) \times SU_R(3)$ Algebra
in the $(3,6) \oplus (6, 3)$ Multiplet} \label{ssect:closure36a}

The vector charge $Q^a = \int d {\bf x}J_{0}^a (t,{\bf x})$ defined
by the current in
\begin{eqnarray}
{\bf J}_{\mu}^{a} &=& \left(\overline{N} \gamma_{\mu} \, {\bf
F}_{(8)}^{a} \,N \right) + \left(\overline{\Delta} \gamma_{\mu} \,
{\bf F}_{(10)}^{a} \,\Delta \right)  \, , \label{e:vec1SU(3)63} \
\end{eqnarray}
together with the axial
charge $Q_{5}^a = \int d {\bf x}J_{05}^a (t,{\bf x})$ defined by the
current in
\begin{eqnarray}
{\bf J}_{\mu 5}^{a} &=& \,\overline{N} \gamma_{\mu} \gamma_{5}
\left({\rm ({\bf D}^{a} + {2\over3} {\bf F}_{(8)}^{a})} N +
\frac{2}{\sqrt{3}} {\rm {\bf T}}^{a} \Delta \right)
+ \, \overline{\Delta} \gamma_{\mu} \gamma_{5}
\left(\frac{2}{\sqrt{3}} {\rm {\bf T}}^{a \dagger} N + \frac{1}{3}
{\rm {\bf F}}_{(10)}^{a} \Delta \right)  \, , \label{e:axi1SU(3)63} \
\end{eqnarray}
ought to close the chiral
algebra defined by Eqs.~(\ref{e:cc3VVV}), (\ref{e:cc3AVA}) and
(\ref{e:cc3AAA}). Eqs.~(\ref{e:cc3VVV}) and (\ref{e:cc3AVA}) hold
here, whereas Eq.~(\ref{e:cc3AAA}) is once again the non-trivial
one: the fractions $\frac23$ and $\frac13$ in the diagonal $F$
charge of $N$ ($Q_{5}^a(N)$) and $\Delta$ axial charges,
respectively, and the diagonal $D$ charge of $N$ ($Q_{5}^a(N)$):
\begin{eqnarray}
Q_{5F}^a(N) &=& \frac{2}{3} \int d {\bf x} \, \left(\overline{N}
\gamma_{0} \gamma_{5} \,{\bf F}_{(8)}^{a} \,N \right)~
\label{e:offdiag1fSU(3)} \, , \\
Q_{5F}^a(\Delta) &=& \frac{1}{3} \int d {\bf x} \,
\left(\overline{\Delta} \gamma_{0} \gamma_{5} \,{\bf F}_{(10)}^{a}
\,\Delta\right) ~ \label{e:offdiag1DeltaSU(3)} \, , \\
Q_{5D}^a(N) &=&  ~~~ \int d {\bf x} \, \left(\overline{N} \gamma_{0}
\gamma_{5} \,{\bf D}^{a} \,N \right)~ \label{e:offdiag1dSU(3)_2} \, ,
\end{eqnarray}
lead to
\begin{eqnarray}
\left[Q_{5D+F}^a(N) , Q_{5D+F}^b(N) \right] &=& \int d {\bf x}
\Bigg(\overline{N} \gamma_{0} \,\Big(\big({\bf D}^{a} + \frac{2}{3}
{\bf F}_{(8)}^{a}\big) \big({\bf D}^{b} + \frac{2}{3} {\bf
F}_{(8)}^{b}\big) \\ \nonumber && - \big({\bf D}^{b} + \frac{2}{3}
{\bf F}_{(8)}^{b}\big) \big({\bf D}^{a} + \frac{2}{3} {\bf
F}_{(8)}^{a}\big) \Big) N \Bigg)
\label{e:ccrAAAdiag1SU(3)} \, , \\
\left[Q_{5F}^a(\Delta) , Q_{5F}^b(\Delta) \right] &=& i f^{abc}
\frac19 Q^c(\Delta) \, , \label{e:ccrAAAdiag2SU(3)}\
\end{eqnarray}
lead to ``only'' part of the $N$ and $\Delta$ vector charges,
respectively, on the right-hand-side of
Eq.~(\ref{e:ccrAAAdiag1SU(3)}) and (\ref{e:ccrAAAdiag2SU(3)}).

Once again, it turns out that the off-diagonal terms in the axial
charge
\begin{eqnarray}
Q_{5}^a(N,\Delta) &=& \int d {\bf x} \, \Bigg( \frac{2}{\sqrt{3}}
\left(\overline{N} \gamma_{0} \gamma_{5} \,{\bf T}_{10/8}^a
\,\Delta + \overline{\Delta} \gamma_{0} \gamma_{5} \,{\bf
T}_{10/8}^{a \dagger} \,N \right) \Bigg)
\label{e:axioffdiag1Sigma2} \, ,
\end{eqnarray}
play a crucial role in the closure of the chiral algebra
Eq.~(\ref{e:cc3AAA}). The additional terms in the commutator add up
to
\begin{eqnarray}
\nonumber && \left[Q_{5}^a(N,\Delta) , Q_{5}^b(N,\Delta) \right]
\\ &=&
\frac{4}{3}\int d {\bf x} \left(\overline{N} \gamma_{0}
\,\left({\bf T}_{10/8}^{a} {\bf T}_{10/8}^{b\dagger} - {\bf
T}_{10/8}^{b} {\bf T}_{10/8}^{a \dagger} \right) \,N +
\overline{\Delta} \gamma_{0} \,\left({\bf T}_{10/8}^{a\dagger}
{\bf T}_{10/8}^{b} - {\bf T}_{10/8}^{b\dagger} {\bf
T}_{10/8}^{a}\right) \Delta \right) \label{e:ccrAAVS36a2} \, ,
\end{eqnarray}
which provide the ``missing'' factors due to the following
properties of the off-diagonal flavor operators ${\bf T}^{i}$ and
${\bf D}^{i}$ matrices
\begin{eqnarray}
i \,f^{ijk} ({\bf F}_{(8)}^{k})  &=& \Big(\big({\bf D}^{i} +
\frac{2}{3} {\bf F}_{(8)}^{i}\big) \big({\bf D}^{j} + \frac{2}{3}
{\bf F}_{(8)}^{j}\big) - \big({\bf D}^{j} + \frac{2}{3} {\bf
F}_{(8)}^{j}\big) \big({\bf D}^{i} + \frac{2}{3} {\bf
F}_{(8)}^{i}\big)\Big)
\\ \nonumber && + {4\over3} ({\bf T}^{i}_{10/8} {\bf
T}^{j\dagger}_{10/8} - {\bf T}^{j}_{10/8} {\bf T}^{i\dagger}_{10/8})
\, , \nonumber \\ i \frac23 \,f^{ijk} {\bf F}_{(10)}^{k} &=& {\bf
T}_{10/8}^{i\dagger} {\bf T}_{10/8}^{j} - {\bf T}_{10/8}^{j\dagger}
{\bf T}_{10/8}^{i} \, . \label{e:app2bSigma}
\end{eqnarray}
Therefore, the chiral algebra Eqs.~(\ref{e:cc3VVV}),
(\ref{e:cc3AVA}) and (\ref{e:cc3AAA}) closes in spite, or perhaps
because of the apparent fractional axial charges ($\frac23$ and
$\frac13$).

\end{document}